\documentclass[aps,prb,twocolumn,showpacs,superscriptaddress]{revtex4-2}
\usepackage{amsmath,amssymb}
\usepackage{graphicx}
\usepackage{wasysym}
\usepackage{amsfonts}
\usepackage{bm}
\usepackage{enumerate}
\usepackage{color}
\usepackage{url}
\usepackage[colorlinks=true,citecolor=blue,urlcolor=blue]{hyperref}
\usepackage{epstopdf}
\usepackage{latexsym}
\usepackage[caption=false,singlelinecheck=false]{subfig}

\newcommand{\bra}[1]{\ensuremath{\langle #1 |}}
\newcommand{\ket}[1]{\ensuremath{| #1 \rangle}}

\newcommand{\bea}{\begin{eqnarray}}
\newcommand{\eea}{\end{eqnarray}}
\newcommand{\be}{\begin{eqnarray}}
\newcommand{\ee}{\end{eqnarray}}
\newcommand{\bw}{\begin{widetext}}
\newcommand{\ew}{\end{widetext}}

\newcommand{\bs}{\boldsymbol}

\begin{document}
	
\title{Shedding Light on Microscopic Details: \\2D Spectroscopy of 1D Quantum Ising Magnets}

\author{GiBaik Sim}
\affiliation{Technical University of Munich, TUM School of Natural Sciences, Physics Department, 85748 Garching, Germany}
\affiliation{Munich Center for Quantum Science and Technology (MCQST), Schellingstr. 4, 80799 M{\"u}nchen, Germany}

\author{Frank Pollmann}
\affiliation{Technical University of Munich, TUM School of Natural Sciences, Physics Department, 85748 Garching, Germany}
\affiliation{Munich Center for Quantum Science and Technology (MCQST), Schellingstr. 4, 80799 M{\"u}nchen, Germany}
	
\author{Johannes Knolle}
\affiliation{Technical University of Munich, TUM School of Natural Sciences, Physics Department, 85748 Garching, Germany}
\affiliation{Munich Center for Quantum Science and Technology (MCQST), Schellingstr. 4, 80799 M{\"u}nchen, Germany}
\affiliation{Blackett Laboratory, Imperial College London, London SW7 2AZ, United Kingdom}

\date{\today}
\begin{abstract}
The identification of microscopic models describing the low-energy properties of correlated materials has been a central goal of spectroscopic measurements. We demonstrate how 2D non-linear spectroscopy can be used to distinguish effective spin models whose linear responses show similar behavior. Motivated by recent experiments on the quasi-1D Ising magnet CoNb$_2$O$_6$, we focus on two proposed models, the ferromagnetic twisted Kitaev chain with bond dependent interactions and the transverse field Ising model. The dynamical spin structure factor probed in linear response displays similar broad spectra for both models from their fermionic domain wall excitations. In sharp contrast, the 2D non-linear spectra of the two models show clear qualitative differences: those of the twisted Kitaev model contain off-diagonal peaks originating from the bond dependent interactions and transitions between different fermion bands absent in the transverse field Ising model. We discuss the different signatures of spin fractionalization in integrable and non-integrable regimes of the models and their connection to experiments.
\end{abstract}

\maketitle
\section{Introduction}
The possibility to understand the microscopics of correlated quantum materials is closely connected to advances in spectroscopic techniques~\cite{devereaux2007inelastic}. In addition to the traditional use of linear response probes, \textit{two-dimensional coherent spectroscopy} (2DCS)~\cite{hamm2011concepts, dorfman2016nonlinear} promises to provide additional information because of its ability to access multi-time correlation functions sensitive to interactions between excitations. Probing the non-linear optical response of the target system, 2DCS has been used to study vibrational and electronic excitations in molecules~\cite{mukamel2000multidimensional} and exciton resonances in quantum wells~\cite{zhang2007polarization,cundiff2013optical}. Additionally, recent advances with terahertz sources put the technique in the proper energy ranges for studying optical excitations of magnetic materials~\cite{lu2017coherent}. Unlike conventional one dimensional (1D) spectroscopy and standard inelastic neutron scattering, 2DCS reveals not only the linear response of spin flips but has more direct access to the interplay of intrinsic excitations of magnets. Along this line, it was theoretically proposed that such interplay can be used to identify the presence of fractionalized particles~\cite{wan2019resolving,li2021photon,choi2020theory,qiang2023probing}, their self-energies~\cite{hart2022self} and the effect of interactions between them~\cite{fava2022divergent,gao2023two}.
\\

In this paper, we show that 2DCS can be a powerful tool for quantifying the microscopic model parameters of quantum magnets. Concretely, we consider 2DCS as a means for distinguishing between two alternative model descriptions of Ising chain magnets, the ferromagnetic twisted Kitaev model (TKM) with bond dependent spin exchange terms and the transverse field Ising model (TFIM). In both cases, spin flip excitations fractionalize into domain wall excitations leading to very similar linear response spectra but distinct qualitative differences in 2DCS. 

Our study is motivated by previous works~\cite{fava2020glide,morris2021duality}, which proposed that the field dependent behavior of CoNb$_2$O$_6$, long believed the best example of an Ising chain magnet~\cite{maartense1977field,kobayashi1999three,coldea2010quantum,morris2014hierarchy,kinross2014evolution}, is in fact well captured by the TKM. We first confirm that the linear response of the TKM and TFIM is indeed similar which complicates the identification of the microscopic description. Second, as our main result, we establish that there are significant differences between the 2D spectra of the TKM and TFIM: (i) the magnetic second order susceptibility $\chi_{xxx}^{(2)}$ vanishes for the TKM due to the presence of a $\hat z$-glide symmetry~\cite{fava2020glide} while it is finite for the TFIM; (ii) the third order susceptibility $\chi_{xxxx}^{(3)}$ contains off-diagonal peaks from inter-band fermion transitions for the TKM which are absent for the TFIM. Taking into account the experimentally relevant canting angle~\cite{kobayashi2000low} between the crystal axis $\hat a$ and local axis $\hat x$ in CoNb$_2$O$_6$, we also compute the easier accessible 2D spectrum, $\chi_{yyy}^{(2)}$, of the TKM. We find that $\chi_{yyy}^{(2)}$ becomes finite and contains off-diagonal signals with an external transverse field along $\hat y$, which breaks the $\hat z$-glide symmetry and the integrability of the model, using infinite matrix-product state (MPS) techniques~\cite{white1992density,vidal2007classical,sim2023nonlinear}. We also confirm that such peaks persist in the presence of additional $XX$-type interactions, which can be relevant in CoNb$_2$O$_6$~\cite{kjall2011bound,robinson2014quasiparticle,fava2020glide}.

Our paper is structured as follows. We first briefly review the TKM and TFIM in Sec.~\ref{sec:model} and confirm the similarity of their linear response spectra in Sec.~\ref{sec:linear}. We then discuss their non-linear response and compare the differences of the 2DCS spectra in Sec.~\ref{sec:nonlinear}. In Sec.~\ref{sec:glide}, we investigate the effect of glide symmetry breaking on the 2DCS spectra of the TKM and discuss the relevance of our findings for CoNb$_2$O$_6$. We conclude with a discussion and outlook in Sec.~\ref{sec:conclude}.

\section{Two microscopic models}
\label{sec:model}
We first introduce the TKM~\cite{morris2021duality} described by the following Hamiltonian
\bea
\nonumber
H_{\text{TKM}} = -J \sum_{i=1}^{L'}\big[ \tilde{\sigma}_{2i-1} (\theta)\tilde{\sigma}_{2i}(\theta)+\tilde{\sigma}_{2i}(-\theta)\tilde{\sigma}_{2i+1}(-\theta) \big].\\
\label{eq:TK}
\eea
Here, $J > 0$ is the ferromagnetic exchange parameter, $L'=L/2$ is the number of unit cells, each containing two sites, and $\tilde{\sigma}_{i}(\theta) \!\equiv\! \cos(\theta)\,\sigma_{i}^z +\sin(\theta)\,\sigma_{i}^y$. Such linear combinations imply that the interaction on each odd (even) bond is characterized by the Ising easy axis with an angle $\pm \theta$~\cite{you2016quantum}. The TKM respects two different glide symmetries, $G_y \equiv T_c e^{(i\pi/2) \sum_i^{L} \sigma^y_{i}} $ and $G_z \equiv T_c e^{(i\pi/2)\sum_i^{L} \sigma^z_{i}} $, where $T_c$ is a translation operator by half a unit cell~\cite{fava2020glide}. When $0<\theta<\pi/4$, the TKM admits a doubly degenerate ferromagnetic ground state, polarized along the easy axis $\hat z$. In this regime, the ground state spontaneously breaks $G_y$, but still preserves one global symmetry $G_z$. Below we fix $\theta\!=\!\pi/12$ for the TKM which is close to the value used in Ref.~\onlinecite{morris2021duality} to describe CoNb$_2$O$_6$. In this case, the elementary excitations of the TKM are domain walls between the two degenerate ground states, similar to the ferromagnetic TFIM with interactions given by
\bea
H_{\text{TFIM}}\!=\! -J \sum_{i=1}^{L}\sigma_i^z\sigma_{i+1}^z \!-\! h_x \sum_{i=1}^{L}\sigma_i^x.
\label{eq:TFIM}
\eea 
Below, we fix $h_x/J\!=\!1/2$ at which the model also stabilizes a doubly degenerate ferromagnetic ground state.

Performing the Jordan-Wigner transformation, which maps the Pauli operators to fermion operators, and Bogoliubov transformation, we can rewrite both the TKM and TFIM as non-interacting fermionic models. The TKM then reads,
\bea
\nonumber
H_{\text{TKM}}&\!=\!& \sum_{k>0} l_k (\alpha^\dagger_k \alpha_k-\alpha_{-k} \alpha_{-k}^\dagger) + \lambda _k (\beta^\dagger_k \beta_k-\beta_{-k} \beta_{-k}^\dagger).\\
\label{eq:TK_JW}
\eea
Here, $\alpha_k$ and $\beta_k$ represent the two different bands with dispersion relations $2l_k$ and $2\lambda_k$ respectively for the TKM in momentum space representation~(See Appendix~\ref{app:JW} for details). The Hamiltonian in Eq.~(\ref{eq:TK_JW}) can be interpreted as a four level system with a momentum pair $\pm k$, where energies of states are $-\lambda_k$, $-l_k$, $l_k$, and $\lambda_k$. In the following, we denote such states by $|0\rangle$, $|1\rangle$, $|2\rangle$, and $|3\rangle$. Thus the TKM corresponds to an ensemble of decoupled four level systems. The TFIM in fermionic formulation reads,
\bea
H_{\text{TFIM}}&\!=\!& \sum_{k>0}  \epsilon_k (\gamma^\dagger_k \gamma_k-\gamma_{-k} \gamma_{-k}^\dagger),
\label{eq:TFIM_JW}
\eea
where $\gamma_k$ represents a single band with dispersion $2\epsilon_k$. The TFIM corresponds to an ensemble of decoupled two level systems with the energy gap $2\epsilon_k$ and is clearly distinct from the TKM.

\section{Linear response structure factor}
\label{sec:linear}
We first briefly compare the linear response, i.e., the dynamical structure factor
\bea
\nonumber
S_ {xx}(k, \omega) = \frac{1}{4} \int \mathrm{d}t \sum_{j} e^{i\omega t - i k (r_j - r_{L/2})} \langle \sigma^x_j(t)\sigma^x_j(0) \rangle\\
\label{equ:dsf}
\eea
of the TKM and TFIM~\cite{laurell2023spin}. In both systems, a spin flip excites a pair of domain walls~(fermions) with net momentum $k$. Such fractionalization of the excitations only yields a broad continuous spectrum. In Fig.~\ref{fig:dsf}, we plot $S_{xx}(k, \omega)$ computed using MPS simulations with open boundary conditions (See Appendix~\ref{app:dfs} for details)~\cite{paeckel2019time}. Remarkably, the two models show a qualitatively similar spectrum, indicating the difficulty of using a conventional probe like inelastic neutron scattering for distinguishing between the two model descriptions.

\begin{figure}[t!]
	\includegraphics[width = \columnwidth]{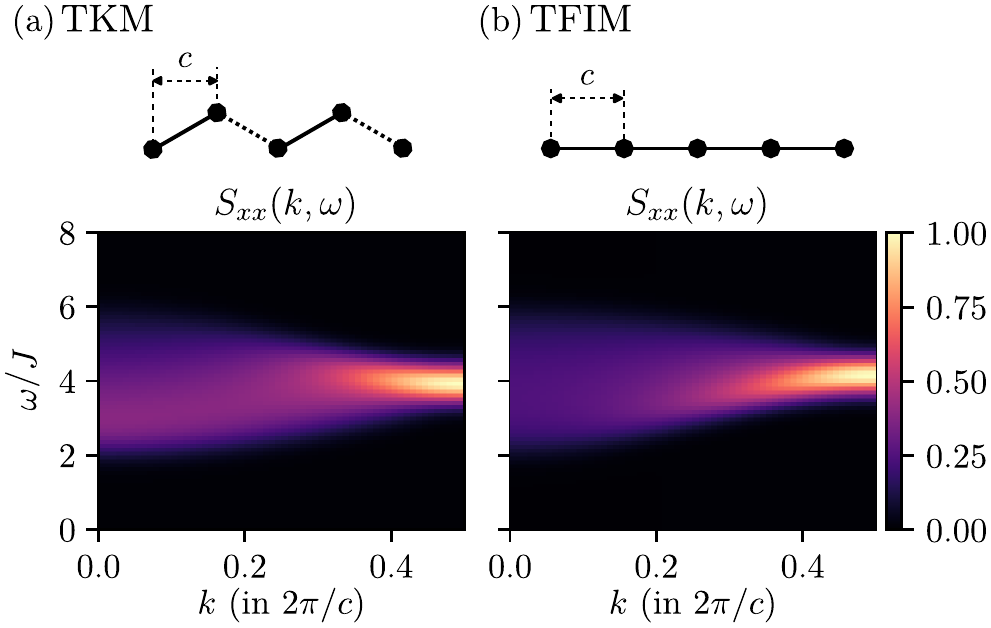}		
	\caption{(color online) (a,b) Dynamical spin structure factor $S_{xx}(k, \omega)$ of the ferromagnetic TKM and TFIM. $k$ and $\omega$ represent the momentum and frequency, respectively. The MPS simulations are done for an open chain using time evolving block decimation method~\cite{vidal2003efficient,white2004real}, which provides an efficient way to perform a real time evolution in 1D spin systems.}
	\label{fig:dsf}
\end{figure}

\section{non-linear response}
\label{sec:nonlinear}

\begin{figure*}[]
	\includegraphics[width=2.05\columnwidth]{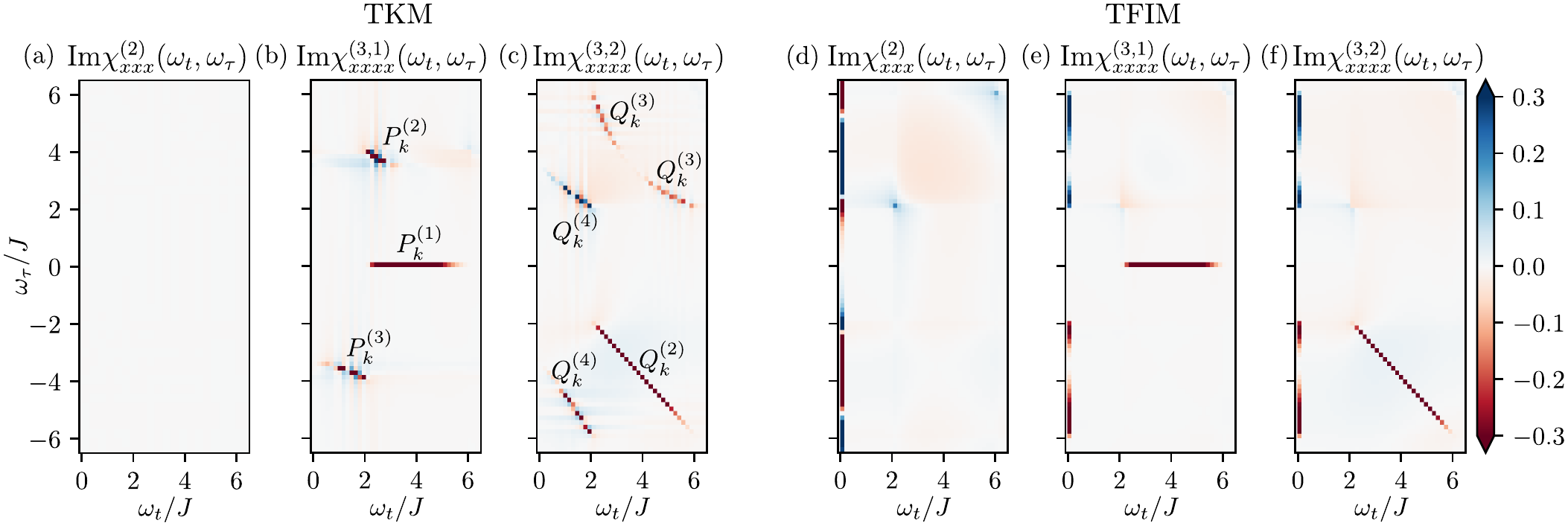}		
	\caption{(color online)(a)-(c) From the left to right: Imaginary part of Fourier transformed $\chi^{(2)}_{xxx}(t, \tau+t)$, $\chi^{(3,1)}_{xxxx}(t, \tau+t,\tau+t)$, and $\chi^{(3,2)}_{xxxx}(t,t,\tau+t)$ of the TKM. (d)-(f) From the left to right: Im$\chi^{(2)}_{xxx}(\omega_t, \omega_\tau)$, Im$\chi^{(3,1)}_{xxxx}(\omega_t, \omega_\tau)$, and Im$\chi^{(3,2)}_{xxxx}(\omega_t, \omega_\tau)$ of the TFIM. The 2D spectra are obtained for a periodic chain of size $L\!=\!220$ over the time range $Jt, J\tau\!=\!40$. Since Im$\chi^{(n)}(\omega_t, \omega_\tau)\!=\!-$Im$\chi^{(n)}(-\omega_t, -\omega_\tau)$, the results are only shown in the first and fourth frequency quadrant. See Appendix~\ref{app:kubo} for the real part. All the data are rescaled such that the maximal absolute value is 1.}
	\label{fig:sig}
\end{figure*}

Next, we focus on the non-linear 2DCS response and introduce a two-pulse setup, as previously considered in Ref.~\onlinecite{wan2019resolving} for the TFIM. In this setup, two magnetic pulses $B_0$ and $B_\tau$ both polarized along $\hat \alpha$ direction,
\bea
\bs B (T) = \mathcal{B}_0 \delta(T) \hat \alpha +  \mathcal{B}_{\tau} \delta(T-\tau) \hat \alpha,
\label{eq:pulse}
\eea
arrive at the target system successively at time $T\!=\!0$ and $T\!=\!\tau$. Here, $\mathcal{B}_{0,\tau}$ are the strength of the pulse over the spatial areas where the pulses $B_{0,\tau}$ reach the system. The two pulses induce magnetization $M_{0,\tau}^{\alpha}(T)$ of the system measured at time $T\!=\!\tau+t$. To subtract the induced magnetization from the linear response, two additional experiments each with a single pulse $B_0$ or $B_\tau$ are performed which measure $M^\alpha_{0}(T)$ or $M^\alpha_{\tau}(T)$, respectively. The non-linear magnetization $M^\alpha_\mathrm{NL}(T) \equiv M^\alpha_{0,\tau}(T)-M^\alpha_{0}(T)-M^\alpha_{\tau}(T)$ can be expanded as
\bea
M^\alpha_\mathrm{NL}(T) &=& \mathcal{B}_0 \mathcal{B}_\tau \chi_{\alpha\alpha\alpha}^{(2)}(t,\tau + t ) \nonumber \\
&+& (\mathcal{B}_0)^2 \mathcal{B}_\tau \chi_{\alpha\alpha\alpha\alpha}^{(3,1)}(t,\tau + t ,\tau + t ) \nonumber \\ 
&+& \mathcal{B}_0 (\mathcal{B}_\tau)^2 \chi_{\alpha\alpha\alpha\alpha}^{(3,2)}(t,t,\tau + t ) + O(B^4)
\label{eq:signal}
\eea
and directly measures the second and higher order magnetic susceptibilities. 

Due to its exact solubulity, we can analytically calculate the $\chi_{xxx}^{(2)}$ and $\chi_{xxxx}^{(3)}$ susceptibilities of the TKM (See Appendix~\ref{app:kubo} for details). A formulation for the TFIM is explicitly given in Ref.~\onlinecite{wan2019resolving}. 

\subsection{2nd order susceptibility}
We start with the second order susceptibility, which is given by
\bea
\nonumber
\chi^{(2)}_{xxx}(t,\tau+t)\!=\!\frac{-\Theta(t)\Theta(\tau)}{L}\langle[[M^x(\tau+t),M^x(\tau)],M^x(0)]\rangle,\\
\eea 
where $M^x(T)\equiv\frac{1}{2}\sum_i e^{iHT} \sigma^x_i e^{-iHT}$ represents the total magnetization along $\hat x$ direction in the Heisenberg picture \cite{wan2019resolving}. To formulate $\chi^{(2)}_{xxx}$ of the TKM, one needs to calculate  expectation values of the form
\bea
\bra{g} M^x(T_1)M^x(T_2)M^x(T_3)\ket{g}.
\label{eq:chi2_general}
\eea
Here, $|g\rangle$ represents the ferromagnetic ground state of the TKM, which is invariant under the $\hat z$-glide operation: $G_z|g\rangle\!=\!|g\rangle$. At the same time, the operator $M^x(T_1)M^x(T_2)M^x(T_3)$ is odd under the same glide operation: $G_z M^x(T_1)M^x(T_2)M^x(T_3) G_z^\dagger \!=\! -M^x(T_1)M^x(T_2)M^x(T_3)$. The invariance of $|g\rangle$ and oddness of the operator $M^x(T_1)M^x(T_2)M^x(T_3)$ under $G_z$ makes Eq.~(\ref{eq:chi2_general}) vanish and $\chi^{(2)}_{xxx}$ is zero [Fig.~\ref{fig:sig}(a)] unless additional symmetry-breaking terms are added to the model. Such a non-linear spectrum of the TKM is clearly distinct from the one of the TFIM [Fig.~\ref{fig:sig}(d)] which contains a strong terahertz rectification signal in $\chi_{xxx}^{(2)}$~\cite{wan2019resolving}. When the second order susceptibility vanishes, the third or higher order susceptibilities dominate the non-linear response of the system.

\subsection{3rd order susceptibility}

We now turn to the next leading order response, i.e., $\chi_{xxxx}^{(3,1)}$ and $\chi_{xxxx}^{(3,2)}$. Using the basis introduced in Eq.~(\ref{eq:TK_JW}), we can also express $M^x(T)$ in terms of fermion operators and obtain the formula for the third order susceptibility of the TKM analytically:
\bea
\nonumber
\chi^{(3,1)}_{xxxx}(t, \tau+t,\tau+t)\!=\!\frac{\Theta(t)\Theta(\tau)}{L}\sum_{k>0} P^{(1)}_{k}\!+\!P^{(2)}_{k}\!+\!P^{(3)}_{k}
\eea
with
\bea
P^{(1)}_k &\!=\!& -8 c_k^4 \big[\sin \big(2 l_k t\big) \!+\! (l_k \leftrightarrow \lambda_k) \big],
\nonumber
\label{eq:P1}
\\
\nonumber
P^{(2)}_k &\!=\!& 8 (c_k^4\!-\!c_k^2) \big[\sin \big( 2l_kt \!+\! (l_k+\lambda_k) \tau \big) \!+\! (l_k \leftrightarrow \lambda_k) \big],
\label{eq:P2}
\\
\nonumber
P^{(3)}_k &\!=\!& 8 (c_k^2\!-\!c_k^4) \big[\sin \big( (l_k\!-\!\lambda_k)t + (l_k\!+\!\lambda_k)\tau \big) \!+\! (l_k \leftrightarrow \lambda_k) \big],
\label{eq:P3}
\eea
where $c_k$ is the matrix element of the magnetization along $\hat x$ direction in the basis introduced in Eq.~(\ref{eq:TK_JW}) (See Appendix~\ref{app:kubo} for  definitions). In $\chi_{xxxx}^{(3,1)}$, $P^{(1-3)}_{k}$ represent different two-time evolution paths of a fermion pair with momenta $\pm k$ excited by the pulses. Employing the four level picture of the fermionic Hamiltonian in momentum space, we can interpret $P^{(1)}_k$ as follows. The second pulse in the two-pulse setup induces transitions between $|1\rangle$ and $|2\rangle$, resulting in an oscillatory signal with frequency $2l_k$ throughout the time interval $t$ between the second pulse and measurement. This signal is encoded in the first term of $P^{(1)}_k$. The term is not oscillatory in $\tau$ and gives rise to a peak at $(\omega_t, \omega_\tau)\!=\!(2l_k, 0)$ in the frequency domain~[Fig.~\ref{fig:sig}(b)]. Interpreting $\omega_t$ as the detecting frequency and $\omega_\tau$ the pumping frequency, the signal can be understood as a pump probe signal. Such signal is also contained in $\chi_{xxxx}^{(3,1)}$ of the TFIM~[Fig.~\ref{fig:sig}(e)]. $P^{(2)}_k$ contains terms which are oscillatory both in $t$ and $\tau$. Such terms produce non-rephasing like signals at $(\omega_t, \omega_\tau)\!=\!(2l_k, l_k+\lambda_k)$ and $(\omega_t, \omega_\tau)\!=\!(2\lambda_k, l_k+\lambda_k)$, giving rise to off-diagonal peaks in the first frequency quadrant as shown in Fig.~\ref{fig:sig}(b). $P^{(3)}_k$ is distinct from $P^{(1,2)}_k$ in that it contains a term where $t$ and $\tau$ come with opposite signs : the dephasing process during $\tau$ is followed by the rephasing process during $t$. This process induces rephasing like signals which appear as off-diagonal peaks in the fourth quadrant, mirroring the energy range of corresponding fermion pairs~[Fig.~\ref{fig:sig}(b)].

Qualitatively different signals are encoded in $\chi_{xxxx}^{(3,2)}$, which is given as
\bea
\nonumber
\chi^{(3,2)}_{xxxx}(t,t,\tau+t)\!=\!\frac{\Theta(t)\Theta(\tau)}{L}\sum_{k>0} Q^{(1)}_{k}\!+\!Q^{(2)}_{k}\!+\!Q^{(3)}_{k}\!+\!Q^{(4)}_k
\eea
with
\bea
\nonumber
Q^{(1)}_k &\!=\!& -4 c_k^2 \big[\sin \big(2 l_k (t+\tau) \big) \!+\! (l_k \leftrightarrow \lambda_k) \big],
\label{eq:Q1}
\\
\nonumber
Q^{(2)}_k &\!=\!& -4 c_k^4 \big[\sin \big(2 l_k (t-\tau)\big) \!+\! (l_k \leftrightarrow \lambda_k) \big],
\label{eq:Q2}
\\
\nonumber
Q^{(3)}_k &\!=\!& 4 (c_k^4-c_k^2) \big[\sin \big(2 \lambda_k t + 2 l_k \tau \big) \!+\! (l_k \leftrightarrow \lambda_k) \big],
\label{eq:Q3}
\\
\nonumber
Q^{(4)}_k &\!=\!& 8 (c_k^2-c_k^4) \big[ \sin \big( (\lambda_k - l_k ) t + 2 \l_k \tau  \big) \!+\! (l_k \leftrightarrow \lambda_k) \big].
\label{eq:Q4}
\eea
The presence of $Q^{(1)}_k$ and $Q^{(2)}_k$ results in the appearance of diagonal peaks in the frequency domain. $Q^{(1)}_k$ is oscillatory in $t + \tau$ and can induce diagonal non-rephasing signals in the first quadrant. $Q^{(2)}_k$ is unique in that $t$ and $\tau$ come with opposite signs but with the same oscillation frequency $2l_k$ or $2\lambda_k$. Unlike other terms, the phase accumulated during $\tau$ is perfectly canceled out during $t$, regardless of the oscillation frequency. This corresponds to the ``spinon echo", which was discovered in Ref.~\onlinecite{wan2019resolving} for the TFIM~[Fig.~\ref{fig:sig}(f)], and results in a diagonal rephasing signal in the fourth quadrant~[Fig.~\ref{fig:sig}(c)]. $Q^{(3)}_k$ produces non-rephasing like signals at $(\omega_t,\omega_\tau)\!=\!(2\lambda_k,2l_k)$ and $(\omega_t,\omega_\tau)\!=\!(2l_k,2\lambda_k)$, giving rise to off-diagonal peaks in the first quadrant~[Fig.~\ref{fig:sig}(c)]. $Q^{(4)}_k$ contains terms that induce strong off-diagonal peaks in the frequency domain, reflecting the energy range of corresponding states~[Fig.~\ref{fig:sig}(c)].

\subsection{Discussion of 2nd and 3rd order susceptibilities}
The 2D spectra of the TKM and the TFIM show qualitative differences. First, $\chi_{xxx}^{(2)}$ of the TKM vanishes due to $\hat z$-glide symmetry while it is finite for the TFIM. In such a situation, $\chi_{xxxx}^{(3)}$ dominates the non-linear response of the system. $\chi_{xxxx}^{(3)}$ of the TKM contains off-diagonal peaks coming from the staggered interactions, in sharp contrast to $\chi_{xxxx}^{(3)}$ of the TFIM. The emergence of such off-diagonal peaks can be used to distinguish the TKM from the TFIM.

\section{Glide symmetry and $\mathbf{\text{CoNb}_2\text{O}_6}$}
\label{sec:glide}

\begin{figure}[]
	\includegraphics[width = 0.9\columnwidth]{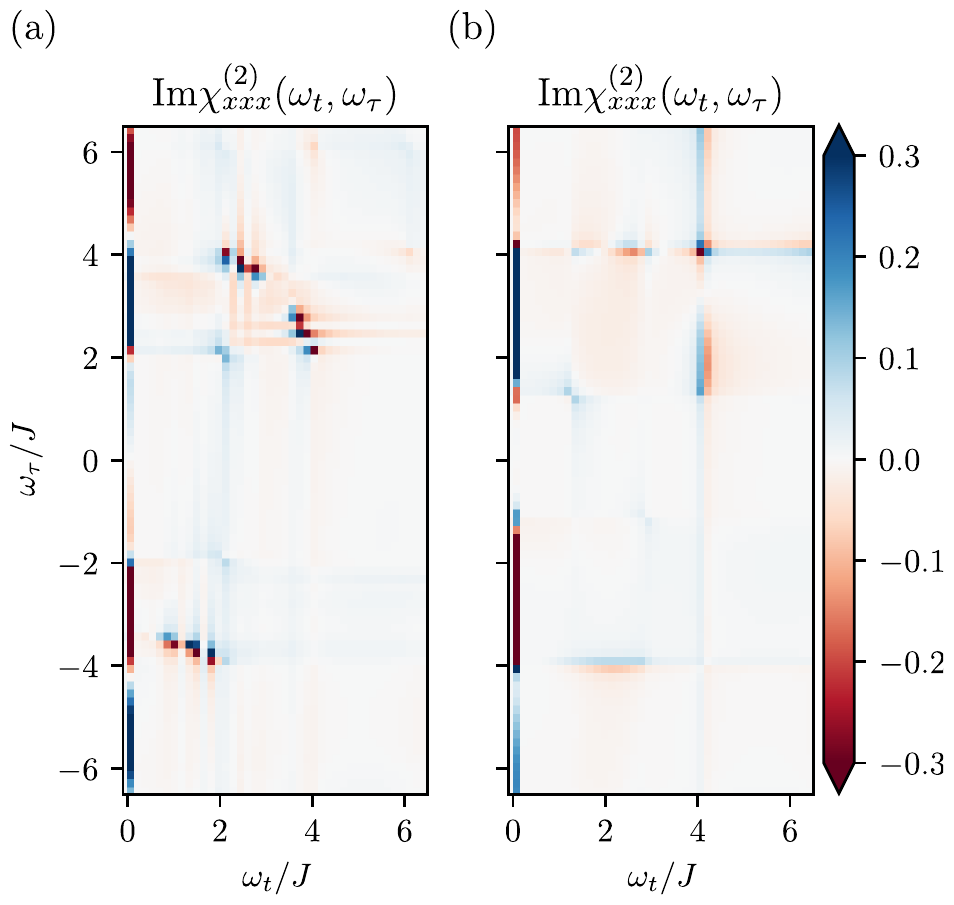}
	\caption{(color online) Im$\chi^{(2)}_{xxx}(\omega_t, \omega_\tau)$ of the TKM with a (a) small transverse field $h_x/J\!=\!1/20$ and (b) strong transverse field $h_x/J\!=\!1/2$. The calculations are done for a periodic chain of $L\!=\!220$ and over the time range $Jt, J\tau \!=\! 40$.}
	\label{fig:chi2_w}
\end{figure}

\subsection{Integrable cases}
We now add a transverse field term $-h_x \sum^L_i \sigma^x_{i}$ to the TKM in Eq.~(\ref{eq:TK}), which breaks the $\hat z$-glide symmetry and makes the leading non-linear susceptibility $\chi_{xxx}^{(2)}$ finite. We then investigate whether off-diagonal peaks arise in $\chi_{xxx}^{(2)}(\omega_t,\omega_\tau)$, revealing the staggered interactions. Note, we only focus on the regime where the ground state remains ferromagnetic. The transverse field term is also quadratic in fermionic operators, allowing for an analytic calculation of non-linear susceptibilities (See Appendix~\ref{app:TKM_t} for details). In Fig.~\ref{fig:chi2_w}(a), we plot Im$\chi_{xxx}^{(2)}(\omega_t,\omega_\tau)$ at a low transverse field $h_x/J\!=\!1/20$. First, it contains a dominant vertical terahertz rectification signal similar to the one of the TFIM~\cite{wan2019resolving}. At the same time, off-diagonal peaks appear, reflecting the energy range of corresponding fermion pairs. In the strong field regime $h_x/J\!=\!1/2$, the amplitude of the off-diagonal peaks becomes relatively weak as shown in Fig.~\ref{fig:chi2_w}(b). It can be understood by the fact that the strength of the staggered terms in the TKM, given by $\pm YZ$-type interactions in Eq.~(\ref{eq:TK}), become relatively small in the strong field limit $h_x/J\to1$ where the full model behaves like the TFIM.

\subsection{Non-integrable cases and $\mathbf{\text{CoNb}_2\text{O}_6}$}

\begin{figure}[t!]
\includegraphics[width = 0.9\columnwidth]{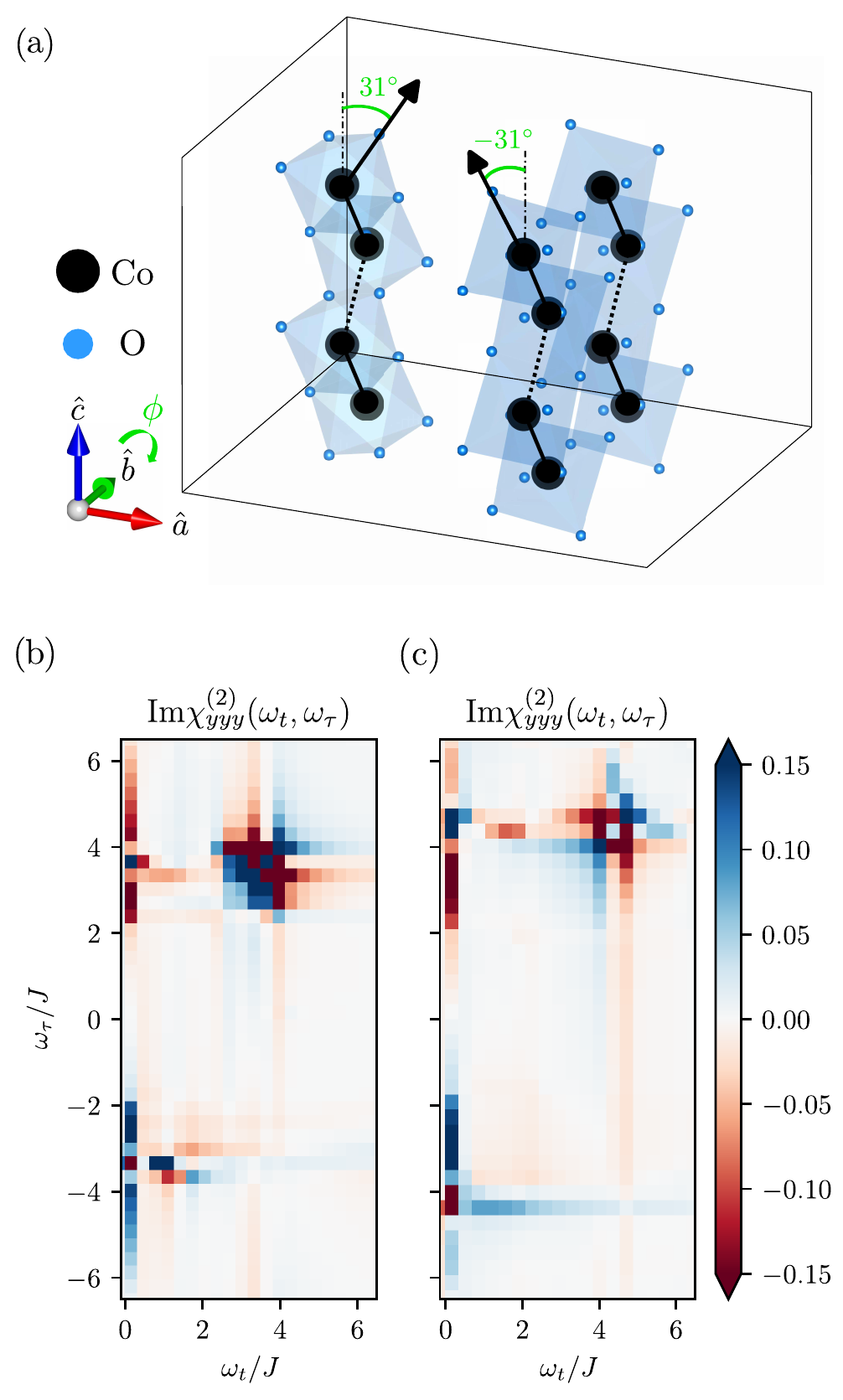}
\caption{(color online) (a) Unit cell of CoNb$_2$O$_6$ where each zigzag chain is aligned along the crystal axis $\hat c$. The black arrows indicate the two different local $\hat z$ axis, which lies in the $\hat a\!-\!\hat c$ plane with the tilting angle $\phi\!=\!\pm31^\circ$~\cite{morris2021duality,kobayashi2000low}. (b) Im$\chi^{(2)}_{yyy}(\omega_t, \omega_\tau)$ of the model in Eq.~(\ref{eq:cno}) with $\theta\!=\!\pi/12$, $J_x/J\!=\!0$, and $h_y/J\!=\!1/20$. (c) Im$\chi^{(2)}_{yyy}(\omega_t, \omega_\tau)$ of the model with $\theta\!=\!\pi/12$, $J_x/J\!=\!1/10$, and $h_y/J\!=\!1/20$.}
\label{fig:chi2_y}
\end{figure}

Our results can be compared to the quasi-1D Ising magnet CoNb$_2$O$_6$, which was recently proposed as a close material realization of the TKM~\cite{fava2020glide,morris2021duality}. In CoNb$_2$O$_6$, cobalt atoms are surrounded by distorted octahedra formed by oxygen atoms. These edge-sharing octahedra form an isolated zigzag 1D chain along the crystal axis $\hat c$, as shown in Fig.~\ref{fig:chi2_y}(a). In this material, the local axis $\hat y$ is exactly aligned to the crystal axis $\hat b$ unlike the local axis $\hat x$, which makes an angle $\phi \!=\! \pm 31^\circ$ with the crystal axis $\hat a$~\cite{morris2021duality,kobayashi2000low}. In this case, $\chi_{yyy}^{(2)}\!=\!\chi_{bbb}^{(2)}$ would be experimentally more pronounced than $\chi_{xxx}^{(2)}$, which is distinct from $\chi_{aaa}^{(2)}$. On the other hand, the accurate model of CoNb$_2$O$_6$ may comprise sub-dominant $XX$-type interactions, which are allowed by the crystal symmetry, as revealed by neutron scattering experiments~\cite{kjall2011bound,robinson2014quasiparticle,fava2020glide}. In this regard, we focus on the model given as
\bea
\nonumber
H = &-&J \sum_{i=1}^{L'}\big[ \tilde{\sigma}_{2i-1} (\theta)\tilde{\sigma}_{2i}(\theta)+\tilde{\sigma}_{2i}(-\theta)\tilde{\sigma}_{2i+1}(-\theta) \big]\\
&-&J_x\sum_i^L \sigma^x_i\sigma^x_{i+1} - h_y \sum^L_i \sigma^y_{i}.
\label{eq:cno}
\eea
which contains the additional $XX$-type interaction  and transverse field term along the $\hat y$ axis. Since the $\hat b$ and $\hat y$ axes are aligned, the transverse term can be included by simply applying an external field along the crystal axis $\hat b$ to CoNb$_2$O$_6$. The model is non-integrable, except in two cases with $\theta\!=\!0$ or $\theta\!\neq\!0$, $J_x\!=\!0$ and $h_y\!=\!0$.

We now calculate $\chi_{yyy}^{(2)}$ of the model in Eq.~(\ref{eq:cno}) in the ferromagnetic regime with fixed $\theta\!=\!\pi/12$, which is close to the value given in Ref.~\cite{morris2021duality} for CoNb$_2$O$_6$, using infinite MPS techniques~\cite{sim2023nonlinear}. The techniques provide a way to calculate the exact $\chi_{yyy}^{(2)}$ for the non-integrable cases and the calculations are done for system sizes $L\!=\!120$ and over the time range $Jt, J\tau\!=\!20$. We checked the dependence of $\chi_{yyy}^{(2)}$ on the bond dimension $\chi$ and the time step $\delta t$, settling on $\chi_{\text{max}}=1000$ and $\delta t\!=\!0.01/J$. We first notice that $\chi_{yyy}^{(2)}$ vanishes unless the $\hat z$-glide symmetry breaking term is finite, $h_y\! \neq \!0$. In Fig.~\ref{fig:chi2_y}(b), we plot Im$\chi_{yyy}^{(2)}(\omega_t,\omega_\tau)$ of the model with $J_x/J\!=\!0$ and $h_y/J\!=\!1/20$. Analogous to Im$\chi_{xxx}^{(2)}(\omega_t,\omega_\tau)$ of the TKM with the transverse field along $\hat x$ direction, it also contains off-diagonal peaks which can signal the presence of the staggered interactions. We also investigate the effect of additional $XX$-type interactions on such off-diagonal peaks regarding CoNb$_2$O$_6$. As shown in Fig.~\ref{fig:chi2_y}(c) for the model with $J_x/J\!=\!1/10$ and $h_y/J\!=\!1/20$, such peaks still appear though the amplitude becomes relatively weak.

\section{Conclusions}
\label{sec:conclude}
In the present work, we propose a way to distinguish two similar models, i.e., the ferromagnetic TKM and TFIM, using 2DCS. In both models, elementary spin flips fractionalize into domain wall excitations, resulting in a qualitatively similar continuum in the linear response dynamical structure factor. In contrast, we show that the 2D non-linear spectrum as a function of $\omega_\tau$ and $\omega_t$, associated with the time interval between a probe and measurement pulse, offers a clear way to discern the two models. Unlike the TFIM, the second order susceptibility $\chi_{xxx}^{(2)}$ vanishes for the TKM due to the presence of a $\hat z$-glide symmetry. Moreover, the third order susceptibility $\chi_{xxxx}^{(3)}$ of the TKM contains non-rephasing and rephasing like signals which appear as off-diagonal peaks in the frequency domain, originating from the presence of bond dependent interactions.

Regarding the canted structure of CoNb$_2$O$_6$, a possible material realization of the TKM, we also investigate the second order susceptibility $\chi_{yyy}^{(2)}$. For the non-integrable regime of the microscopic model we have emloyed the infinite MPS method for calculating the non-linear response. First, we find that $\chi_{yyy}^{(2)}$ of the TKM vanishes unless additional $\hat z$-glide symmetry breaking terms are included. Second, we observe the emergence of off-diagonal peaks with an external transverse field along $\hat y$ axis. Such peaks persist with additional $XX$-type interactions, which can be sub-dominant in CoNb$_2$O$_6$. We expect our results will shed light on the unambiguous identification of the correct microscopic description of CoNb$_2$O$_6$.

Advances in spectroscopic methods, accessible frequency ranges, and improved energy resolution allow for a more precise understanding of correlated quantum systems. We expect that 2DCS is an excellent tool for determining the microscopic parameters of quantum magnets, not only for the one-dimensional example considered here but also for two- and three-dimensional frustrated magnets. 

\begin{acknowledgments} 
We thank R. Coldea, N. P. Armitage, M. Drescher, H.-K. Jin, W. Choi, N.P. Perkins and S. Gopalakrishnan for insightful discussions related to this work. G.B.S. thanks P. d'Ornellas for detailed comments on the manuscript. G.B.S. is funded by the European Research Council (ERC) under the European Unions Horizon 2020 research and innovation program (grant agreement No. 771537). F.P. acknowledges the support of the Deutsche Forschungsgemeinschaft (DFG, German Research Foundation) under Germany’s Excellence Strategy EXC-2111-390814868. J. K. acknowledges support from the Imperial-TUM flagship partnership. The research is part of the Munich Quantum Valley, which is supported by the Bavarian state government with funds from the Hightech Agenda Bayern Plus. Tensor network calculations were performed using the TeNPy Library~\cite{hauschild2018efficient}. Data analysis and simulation codes are available on Zenodo upon reasonable request~\cite{sim_zenodo}.
\end{acknowledgments}


\appendix

\section{Jordan-Wigner formalism}
\label{app:JW}
In this appendix, we introduce the Jordan-Wigner formulation of the TKM. The TKM is written as
\bea
\nonumber
H_{\text{TKM}} = -J \sum_{i=1}^{L'}\big( \tilde{\sigma}_{2i-1}(\theta)\tilde{\sigma}_{2i}(\theta)+\tilde{\sigma}_{2i}(-\theta)\tilde{\sigma}_{2i+1}(-\theta) \big).
\label{eq:TK_s}
\eea
We now introduce the Jordan-Wigner transformation which maps the Pauli operators to fermionic operators through the relations~\cite{you2016quantum}:
\bea
\nonumber
\sigma _{j}^{x} &=& 1-2c_{j}^{\dagger }c_{j}, \\
\nonumber
\sigma _{j}^{y} &=& -i(c_{j}^{\dagger}-c_{j}) \prod_{i<j}\,(1-2c_{i}^{\dagger }c_{i}^{}), \\
\sigma _{j}^{z} &=& (c_{j}^{\dagger}+c_{j}) \prod_{i<j}\,(1-2c_{i}^{\dagger }c_{i}^{}).
\label{eq:JW_s}
\eea
Using Eq.~(\ref{eq:JW_s}), we arrive at a bilinear form in terms of spinless fermions:
\bea
\nonumber
H_{\text{TKM}}&=&\sum_{i=1}^{L'} \Big[ e^{i\theta} c_{2i-1}^{\dagger} c_{2i}^{\dagger} + c_{2i-1}^{\dagger} c_{2i}^{} \\
&+& e^{-i\theta} c_{2i}^{\dagger} c_{2i+1}^{\dagger} + c_{2i}^{\dagger} c_{2i+1}^{} + h.c. \Big].
\eea
We then adopt the Fourier transformation $c_{2j-1}\!=\!\frac{1}{\sqrt{L'}}\sum_{k}e^{-ik j}a_{k},
c_{2j}\!=\!\frac{1}{\sqrt{L'}}\sum_{k}e^{-ik j}b_{k}$ with the discrete momenta $k\!=\!\frac{n\pi}{ L'}, n\!=\! -(L'\!-1), \ldots, (L'\!-3), (L'\! -1)$.
The TKM now takes the form as
\bea
\nonumber
H_{\text{TKM}}&=& \sum_{k} [ B_k^{} a_{k}^{\dagger} b_{-k}^{\dagger}+ A_k^{} a_{k}^{\dagger} b_{k}^{} - A_k^* a_{k}^{}b_{k}^{\dagger}-B_k^* a_{k}^{}b_{-k}]\\
\label{eq:TKM_bdg}
\eea
where $A_k= 1 + e^{ik}$ and $B_k = e^{i\theta} - e^{i(k-\theta)}$. To diagonalize the Hamiltonian Eq.~(\ref{eq:TKM_bdg}), we write it in a matrix form as
\bea
H_{\text{TKM}}=\sum_{k>0}(a^\dagger_k,a_{-k},b^\dagger_k,b_{-k}) \hat{M}_k
\begin{pmatrix}
	a_k \\
	a^\dagger_{-k}\\
	b_k \\
	b^\dagger_{-k}\\
\end{pmatrix}
\label{eq:TKM_bdg_m}
\eea
where
\bea
\nonumber
\hat{M}_k\!=\!\left(\!
\begin{array}{cccc}
	0 & 0 &   S_k   &  P_k+Q_k  \\
	0 & 0 & P_k- Q_k   &    -S_k  \\
	S_k^*  &  P_k^*-Q_k^*   & 0 & 0 \\
	P_k^*+Q_k^*   &  -S_k^*   & 0 & 0
\end{array}\!\right)
\eea
with $P_k=-i ( e^{ik} + 1)\sin\theta, Q_k= ( e^{ik} - 1)\cos\theta$, and $S_k =1+ e^{ik}$. The diagonalization of Eq.~(\ref{eq:TKM_bdg_m}) is achieved by the Bogoliubov transformation,
\bea
(\alpha_k^\dagger, \alpha_{-k}, \beta_k^\dagger, \beta_{-k})~\hat{U}_{k} = (a_k^\dagger, a_{-k}, b_k^\dagger, b_{-k}).
\label{eq:2DXXZ_RDM}
\eea
The Hamiltonian is now diagonalized in the new basis as
\bea
\nonumber
H_{\text{TKM}} = \sum_{k>0} \big[ l_k (\alpha^\dagger_k \alpha_k-\alpha_{-k} \alpha_{-k}^\dagger) + \lambda _k (\beta^\dagger_k \beta_k-\beta_{-k} \beta_{-k}^\dagger) \big]\\
\eea
where $l_k\!=\! \sqrt{\xi_k - \sqrt{\xi_k^2-\tau_k^2}}, ~\lambda_k\!=\!\sqrt{\xi_k + \sqrt{\xi_k^2-\tau_k^2}}$ with $\xi_k=\vert P_k \vert^2 + \vert Q_k \vert^2 +  \vert S_k \vert^2$ and $\tau_k=\vert  P_k^2 - Q_k^2  + S_k^2 \vert$.

\section{Detail of MPS simulation for the dynamical structure factor}
\label{app:dfs}
In this appendix, we provide details of the MPS simulations for the dynamical structure factor~\cite{paeckel2019time}
\small
\bea
\nonumber
&&S_{xx}(k, \omega)\!=\!\frac{1}{4} \int \mathrm{d}t \sum_{j} e^{i\omega t - i k (r_j - r_{L/2})} \langle \sigma^x_j(t)\sigma^x_{L/2}(0) \rangle \\
\nonumber
&&\!=\!\frac{1}{4}\int \mathrm{d}t \sum_{j} \big[e^{i\omega t - i k (r_j - r_{L/2})} e^{i E_g t} \bra g \sigma^{x}_j e^{-iHt} \sigma^{x}_{L/2} \ket g G(t) \big].
\label{equ:a_dsf}
\eea
\normalsize
\begin{enumerate}
\item
Find an MPS approximation of the ground state $\ket{g}$ with an energy $E_g$ using the density matrix renormalization group.
\item
Apply a local operator $\sigma^x_{L/2}$ and obtain $\sigma^x_{L/2}\ket{g}$.
\item
Perform a real time evolution following the local quench $\sigma^x_{L/2}$ using time evolving block decimation method~\cite{vidal2003efficient,vidal2004efficient} to get an MPS which represents $e^{-iHt} \sigma^{x}_{L/2} \ket g$.
\item
Evaluate an overlap of two MPS ``bra" and ``ket" to obtain $\bra g \sigma^{x}_j e^{-iHt} \sigma^{x}_{L/2} \ket g$.
\item
Multiply $e^{iE_gt}$ and $\bra g \sigma^{x}_j e^{-iHt} \sigma^{x}_{L/2} \ket g$.
\item
Apply a discrete Fourier transformation in space that yields the momentum-resolved time-dependent data $S_{xx}(k, t)$.
\item
Perform a Fourier transformation of the time series convoluted with a Gaussian window function $G(t)=e^{-t^2/2\sigma^2}$ to prevent Gibb's oscillations~\cite{white2008spectral,verresen2019avoided,fava2020glide}.
\end{enumerate}
For the result given in Fig.~\ref{fig:dsf}, we set the system size $L=120$, time step size $\delta t=0.02J$, total simulation time $t_{max}=60J$, maximum bond dimension $\chi_{max}=500$, and the Gaussian envelope parameter $\sigma=0.05$.

\section{Magnetic Susceptibilities of the TKM}
\label{app:kubo}
Here, we analytically formulate the linear and non-linear magnetic susceptibilities of the TKM. $M^x$, the total magnetization of the target system along $\hat{x}$ direction, in fermionic formulation reads
\bea
\nonumber
M^x &=& \frac{1}{2}\sum_{i}^{L'}(\sigma^x_{2i-1} + \sigma^x_{2i}) = \sum_{k>0}m^x_k \\
\nonumber
&=& \sum_{k>0} \big[ -a^\dagger_k a_k + a_{-k}a^\dagger_{-k} - b^\dagger_k b_k + b_{-k}b^\dagger_{-k} \big].\\
\eea
$M^x$ can be rewritten in the basis introduced in Eq.~(\ref{eq:2DXXZ_RDM}):
\begin{widetext}
\bea
M^x &=& \sum_{k>0} \big[ -a^\dagger_k a_k + a_{-k}a^\dagger_{-k} - b^\dagger_k b_k + b_{-k}b^\dagger_{-k} \big]
= (a^\dagger_k,a_{-k},b^\dagger_k,b_{-k}) 
\begin{pmatrix}
	-1 & 0 & 0 & 0\\
	0 & 1 & 0 & 0\\
	0 & 0 & -1 & 0\\
	0 & 0 & 0 & 1\\
\end{pmatrix}
\begin{pmatrix}
	a_k \\
	a^\dagger_{-k}\\
	b_k \\
	b^\dagger_{-k}\\
\end{pmatrix}
\nonumber\\
&=& (\alpha^\dagger_k,\alpha_{-k},\beta^\dagger_k,\beta_{-k})\hat{U}_k
\begin{pmatrix}
	-1 & 0 & 0 & 0\\
	0 & 1 & 0 & 0\\
	0 & 0 & -1 & 0\\
	0 & 0 & 0 & 1\\
\end{pmatrix}
\hat{U}_k^{\dagger}
\begin{pmatrix}
	\alpha_k \\
	\alpha^\dagger_{-k}\\
	\beta_k \\
	\beta^\dagger_{-k}\\
\end{pmatrix}
\nonumber\\
&=& (\alpha^\dagger_k,\alpha_{-k},\beta^\dagger_k,\beta_{-k})
\begin{pmatrix}
	0 & c_k & \sqrt{1-c_k^2} & 0\\
	c_k & 0 & 0 & \sqrt{1-c_k^2}\\
	\sqrt{1-c_k^2} & 0 & 0 & -c_k\\
	0 & \sqrt{1-c_k^2} & -c_k & 0\\
\end{pmatrix}
\begin{pmatrix}
	\alpha_k \\
	\alpha^\dagger_{-k}\\
	\beta_k \\
	\beta^\dagger_{-k}\\
\end{pmatrix}.
\label{eq:mag}
\eea
In the Heisenberg picture,
\small
\bea
\nonumber
M^x(t) = \sum_k (\alpha^\dagger_k,\alpha_{-k},\beta^\dagger_k,\beta_{-k}) \left(
\begin{array}{cccc}
	0 & c_k e^{-2 i \lambda_k  t} & \sqrt{1-c_k^2} e^{i t (l_k-\lambda_k )} & 0 \\
	c_k e^{2 i \lambda_k  t} & 0 & 0 & \sqrt{1-c_k^2} e^{-i t (l_k-\lambda_k )} \\
	\sqrt{1-c_k^2} e^{-i t (l_k-\lambda_k )} & 0 & 0 & -c_k e^{-2 i l_k t} \\
	0 & \sqrt{1-c_k^2} e^{i t (l_k-\lambda_k )} & -c_k e^{2 i l_k t} & 0 \\
\end{array}
\right)
\begin{pmatrix}
	\alpha_k \\
	\alpha^\dagger_{-k}\\
	\beta_k \\
	\beta^\dagger_{-k}\\
\end{pmatrix}.\\
\label{eq:hei}
\eea
\end{widetext}
\normalsize

\begin{figure*}[]
\includegraphics[width = 1.5\columnwidth]{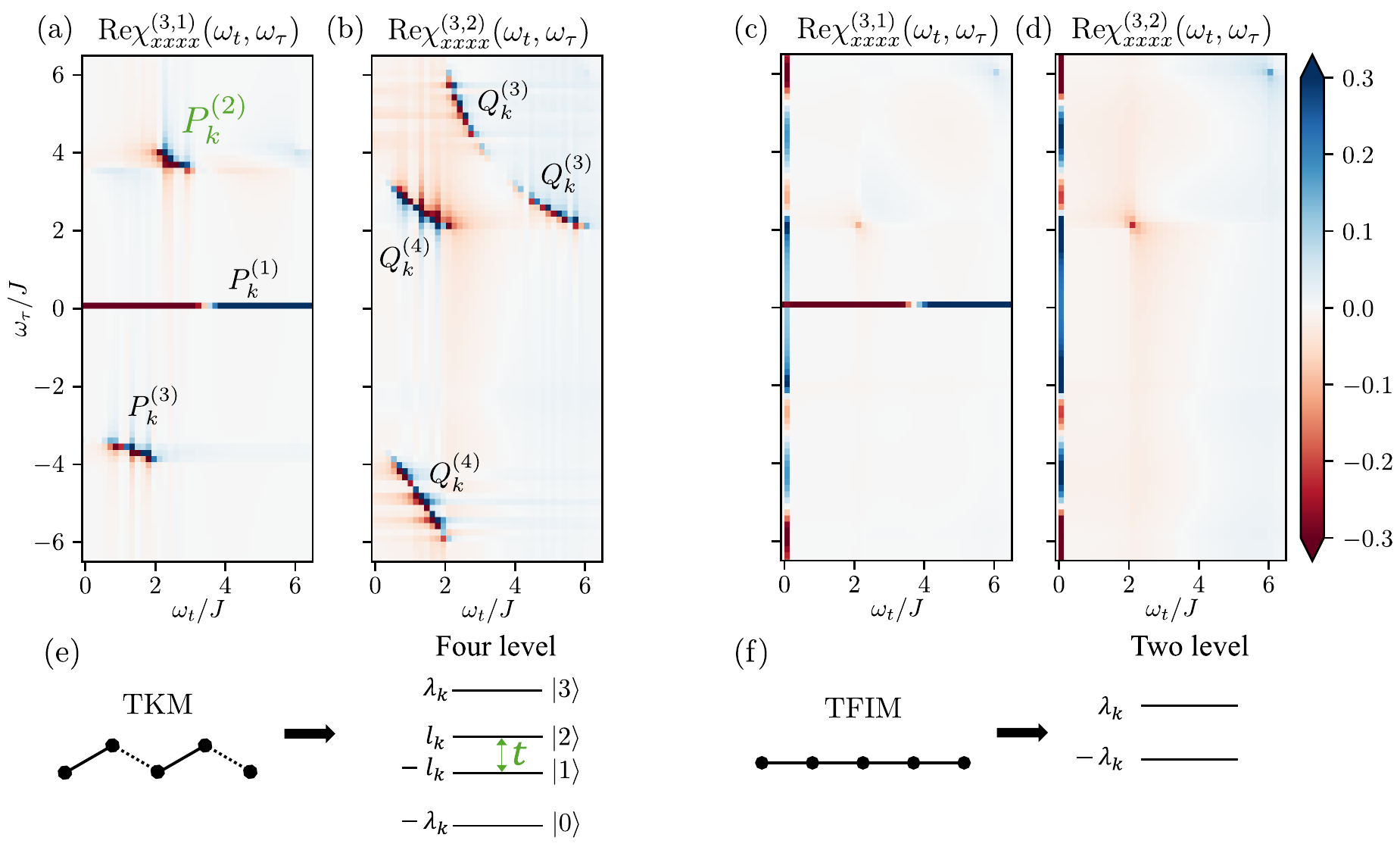}
\caption{(color online) (a,b) Real part of Fourier transformed $\chi^{(3,1)}_{xxxx}(t, \tau+t,\tau+t)$ and $\chi_{xxxx}^{(3,2)}(t, t ,\tau + t )$ of the TKM (four level system). (c,d) Real part of Fourier transformed $\chi^{(3,1)}_{xxxx}(t, \tau+t,\tau+t)$ and $\chi_{xxxx}^{(3,2)}(t, t ,\tau + t )$ of the TFIM (two level system). Since Re$\chi^{(2)}_{xxx}(\omega_t, \omega_\tau)$ $=$ Re$\chi^{(2)}_{xxx}(-\omega_t, -\omega_\tau)$, the result is only shown in first and fourth frequency quadrant.}
\label{fig:c3_x_s}
\end{figure*}

We now calculate the linear and non-linear magnetic susceptibilities of the TKM. We first consider the linear susceptibility $\chi^{(1)}_{xx}(t)$. The starting point is the Kubo formula:
\begin{align}
	\chi^{(1)}_{xx}(t) &= \frac{i \Theta(t)}{L}\langle [M^x(t),M^x(0)] \rangle 
	\nonumber\\
	&= \frac{i \Theta(t)}{L}\sum_{k>0} \langle [m^x_k(t),m^x_k(0)] \rangle 
	\nonumber\\
	&= \frac{2}{L}\sum_{k>0} c_k^2 \big(\sin (2 l_k t)+\sin (2 \lambda_k  t) \big)
	\label{eq:chi1_TK}
\end{align}
where $\langle\cdots\rangle$ represents the average in the ground state. The second equality comes from the fact that $m^x_k$ with different $k$ commute. The second order non-linear susceptibility is given as
\small
\bea
\nonumber
\chi^{(2)}_{xxx}(t,\tau+t) &\!=\!& \frac{i^2\Theta(t)\Theta(\tau)}{L}\langle[[M^x(\tau+t),M^x(\tau)],M^x(0)]\rangle \\
\nonumber
&\!=\!& \frac{i^2\Theta(t)\Theta(\tau)}{L} \sum_{k>0} \langle [[m^x_k (\tau+t),m^x_k (\tau)],m^x_k (0)] \\
&\!=\!&0.
\label{eq:chi2_TK}
\eea
\normalsize
As pointed out in the main text, the second order susceptibility of the TKM vanishes. The formula for the third order non-linear susceptibility is given as
\begin{widetext}
\bea
\chi^{(3)}_{xxxx}(t_3,t_2+t_3,t_1+t_2+t_3) &\!=\!& \frac{i^3\Theta(t_1)\Theta(t_2)\Theta(t_3)}{L}\langle[[[M^x(t_1+t_2+t_3),M^x(t_1+t_2)],M^x(t_1)],M^x(0)]\rangle
\nonumber \\
&\!=\!& \frac{i^3\Theta(t_1)\Theta(t_2)\Theta(t_3)}{L} \sum_{k>0} \langle [[[m^x_k (t_1+t_2+t_3),m^x_k (t_1+t_2)],m^x_k (t_1)],m^x_k(0)] \rangle.
\label{eq:chi3_TK}
\eea
\end{widetext}
We focus on the two limit which correspond $\chi^{(3)}_{xxxx}$ measured in the two-pulse setup, $\chi^{(3,1)}_{xxxx}(t,\tau+t,\tau+t)$ with $t_1 \!\to\! 0, t_2 \!\to\! \tau, t_3 \!\to\! t$ and $\chi^{(3,2)}_{xxxx}(t,t,t+\tau)$ with $t_1 \!\to\! \tau,t_2 \!\to\! 0,t_3 \!\to\! t$:
\bea
\nonumber
\chi^{(3,1)}_{xxxx}(t, \tau+t,\tau+t)\!=\!\frac{\Theta(t)\Theta(\tau)}{L}\sum_{k>0} P^{(1)}_{k}\!+\!P^{(2)}_{k}\!+\!P^{(3)}_{k}
\eea
with
\bea
P^{(1)}_k &\!=\!& -8 c_k^4 \big[\sin \big(2 l_k t\big) \!+\! (l_k \leftrightarrow \lambda_k) \big],
\nonumber
\label{eq:a_P1}
\\
\nonumber
P^{(2)}_k &\!=\!& 8 (c_k^4\!-\!c_k^2) \big[\sin \big( 2l_kt \!+\! (l_k+\lambda_k) \tau \big) \!+\! (l_k \leftrightarrow \lambda_k) \big],
\label{eq:a_P2}
\\
\nonumber
P^{(3)}_k &\!=\!& 8 (c_k^2\!-\!c_k^4) \big[\sin \big( (l_k\!-\!\lambda_k)t + (l_k\!+\!\lambda_k)\tau \big) \!+\! (l_k \leftrightarrow \lambda_k) \big],
\label{eq:a_P3}
\eea
and
\bea
\nonumber
\chi^{(3,2)}_{xxxx}(t,t,\tau+t)\!=\!\frac{\Theta(t)\Theta(\tau)}{L}\sum_{k>0} Q^{(1)}_{k}\!+\!Q^{(2)}_{k}\!+\!Q^{(3)}_{k}\!+\!Q^{(4)}_k
\eea
with
\bea
\nonumber
Q^{(1)}_k &\!=\!& -4 c_k^2 \big[\sin \big(2 l_k (t+\tau) \big) \!+\! (l_k \leftrightarrow \lambda_k) \big],
\label{eq:a_Q1}
\\
\nonumber
Q^{(2)}_k &\!=\!& -4 c_k^4 \big[\sin \big(2 l_k (t-\tau)\big) \!+\! (l_k \leftrightarrow \lambda_k) \big],
\label{eq:a_Q2}
\\
\nonumber
Q^{(3)}_k &\!=\!& 4 (c_k^4-c_k^2) \big[\sin \big(2 \lambda_k t + 2 l_k \tau \big) \!+\! (l_k \leftrightarrow \lambda_k) \big],
\label{eq:a_Q3}
\\
\nonumber
Q^{(4)}_k &\!=\!& 8 (c_k^2-c_k^4) \big[ \sin \big( (\lambda_k - l_k ) t + 2 \l_k \tau  \big) \!+\! (l_k \leftrightarrow \lambda_k) \big],
\label{eq:a_Q4}
\eea
where $c_k$ is the matrix element of the magnetization as given in Eq.~(\ref{eq:mag}).

In Fig.~\ref{fig:c3_x_s}, we plot the real part of Fourier transformed $\chi^{(3,1)}_{xxxx}(t, \tau+t,\tau+t)$ and $\chi_{xxxx}^{(3,2)}(t, t ,\tau + t )$ for the TKM and TFIM. The formulation for the TFIM is explicitly given in Ref.~\onlinecite{wan2019resolving}. $\chi^{(3)}_{xxxx}$ of the TKM (Figs.~\ref{fig:c3_x_s}(a,b)) contains the off-diagonal signals unlike $\chi^{(3)}_{xxxx}$ of the TFIM (Figs.~\ref{fig:c3_x_s}(c,d)). Such signals come from the transition between different excited states, whose presence originates from the bond dependent spin exchange interactions. For example, $P^{(2)}_k$ contains a transition between the first excited state $|1\rangle$ and the second excited state $|2\rangle$ as illustrated in Fig.~\ref{fig:c3_x_s}(e), resulting in an oscillatory signal with frequency $2l_k$ throughout the time interval $t$.

\section{Second order susceptibilities \\ of the TKM with transverse field}
\label{app:TKM_t}
\begin{figure}[b!]
	\includegraphics[width = 0.85\columnwidth]{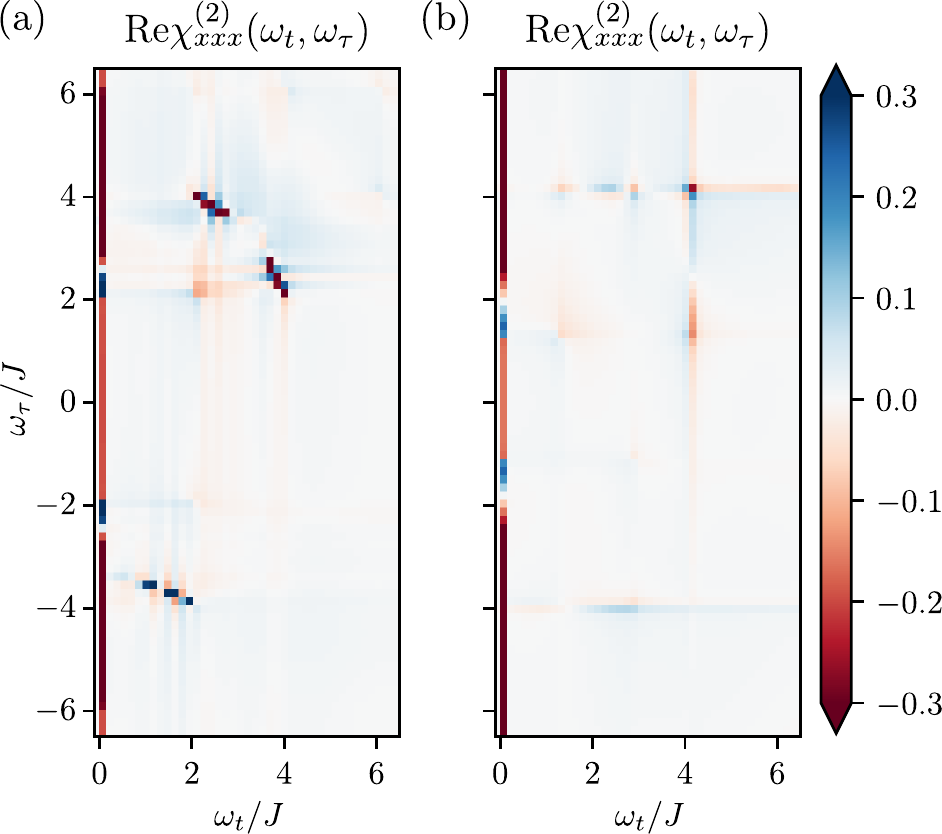}
	\caption{(color online) Real part of Fourier transformed $\chi^{(2)}_{xxx}(t, \tau+t)$ of the TKM with a (a) small transverse field $h_x/J=1/20$ and (b) strong transverse field $h_x/J=1/2$.}
	\label{fig:c2_x_s}
\end{figure}

The TKM with a transverse field along $\hat x$ direction is written as
\bea
\nonumber
H \!=\! &-&J \sum_{i=1}^{L'}\big( \tilde{\sigma}_{2i-1}(\theta)\tilde{\sigma}_{2i}(\theta)\!+\!\tilde{\sigma}_{2i}(-\theta)\tilde{\sigma}_{2i+1}(-\theta)\big) \\
&\!-\!& h_x \sum_{i=1}^{L}\sigma_i^x.
\label{eq:ETK_s}
\eea
We now rewrite the model in terms of spinless fermions using the Jordan-Wigner transformation:
\bea
H_{\text{TKM}}=\sum_{k>0}(a^\dagger_k,a_{-k},b^\dagger_k,b_{-k}) \hat{M}_k
\begin{pmatrix}
	a_k \\
	a^\dagger_{-k}\\
	b_k \\
	b^\dagger_{-k}\\
\end{pmatrix}
\label{eq:ETK_s_f}
\eea
where
\bea
\nonumber
\hat{M}_k\!=\!\left(\!
\begin{array}{cccc}
	2h_x & 0 &   S_k   &  P_k+Q_k  \\
	0 & -2h_x & P_k- Q_k   &    -S_k  \\
	S_k^*  &  P_k^*-Q_k^*   & 2h_x & 0 \\
	P_k^*+Q_k^*   &  -S_k^*   & 0 & -2h_x
\end{array}\!\right)
\eea
with $P_k=-i ( e^{ik} + 1)\sin\theta, Q_k= ( e^{ik} - 1)\cos\theta$, and $S_k =1+ e^{ik}$.
Using the Bogoliubov transformation,
\bea
(\gamma_k^\dagger, \gamma_{-k}, \eta_k^\dagger, \eta_{-k})~\hat{U}_{k} = (a_k^\dagger, a_{-k}, b_k^\dagger, b_{-k}),
\label{eq:2DXXZ_RDM_s}
\eea
Eq.~(\ref{eq:ETK_s_f}) can be diagonalized as
\small
\bea
\nonumber
H_{\text{TKM}} = \sum_{k>0} \big[ l_k (\gamma^\dagger_k \gamma_k-\gamma_{-k} \gamma_{-k}^\dagger) + \lambda_k (\eta^\dagger_k \eta_k-\eta_{-k} \eta_{-k}^\dagger) \big]\\
\eea
\normalsize
where $l_k\!=\! \sqrt{\xi_k + 4 h_x^2 - \sqrt{\xi_k^2-\tau_k^2+16h_x^2|S_k|^2}}, ~\lambda_k\!=\! \sqrt{\xi_k + 4 h_x^2 + \sqrt{\xi_k^2-\tau_k^2+16h_x^2|S_k|^2}}$ with $\xi_k=\vert P_k \vert^2 + \vert Q_k \vert^2 +  \vert S_k \vert^2 + 4 h_x^2$ and $\tau_k=|  P_k^2 - Q_k^2  + S_k^2| + 4 h_x^2$. Then, one can use the Kubo formula given in Appendix~\ref{app:kubo} to obtain non-linear susceptibilities. In Fig.~\ref{fig:c2_x_s}, we plot the real part of Fourier transformed $\chi^{(2)}_{xxx}(t,\tau+t)$ with a low $h_x/J=1/20$ and strong $h_x/J=1/2$ transverse field. As pointed out in the main text, the additional transverse field terms break the $\hat z$-glide symmetry of the TKM and make $\chi^{(2)}_{xxx}$ finite with off-diagonal signals.
\bibliographystyle{apsrev4-2}
\bibliography{tkm_bib}

\begin{thebibliography}{37}%
\makeatletter
\providecommand \@ifxundefined [1]{%
 \@ifx{#1\undefined}
}%
\providecommand \@ifnum [1]{%
 \ifnum #1\expandafter \@firstoftwo
 \else \expandafter \@secondoftwo
 \fi
}%
\providecommand \@ifx [1]{%
 \ifx #1\expandafter \@firstoftwo
 \else \expandafter \@secondoftwo
 \fi
}%
\providecommand \natexlab [1]{#1}%
\providecommand \enquote  [1]{``#1''}%
\providecommand \bibnamefont  [1]{#1}%
\providecommand \bibfnamefont [1]{#1}%
\providecommand \citenamefont [1]{#1}%
\providecommand \href@noop [0]{\@secondoftwo}%
\providecommand \href [0]{\begingroup \@sanitize@url \@href}%
\providecommand \@href[1]{\@@startlink{#1}\@@href}%
\providecommand \@@href[1]{\endgroup#1\@@endlink}%
\providecommand \@sanitize@url [0]{\catcode `\\12\catcode `\$12\catcode
  `\&12\catcode `\#12\catcode `\^12\catcode `\_12\catcode `\%12\relax}%
\providecommand \@@startlink[1]{}%
\providecommand \@@endlink[0]{}%
\providecommand \url  [0]{\begingroup\@sanitize@url \@url }%
\providecommand \@url [1]{\endgroup\@href {#1}{\urlprefix }}%
\providecommand \urlprefix  [0]{URL }%
\providecommand \Eprint [0]{\href }%
\providecommand \doibase [0]{https://doi.org/}%
\providecommand \selectlanguage [0]{\@gobble}%
\providecommand \bibinfo  [0]{\@secondoftwo}%
\providecommand \bibfield  [0]{\@secondoftwo}%
\providecommand \translation [1]{[#1]}%
\providecommand \BibitemOpen [0]{}%
\providecommand \bibitemStop [0]{}%
\providecommand \bibitemNoStop [0]{.\EOS\space}%
\providecommand \EOS [0]{\spacefactor3000\relax}%
\providecommand \BibitemShut  [1]{\csname bibitem#1\endcsname}%
\let\auto@bib@innerbib\@empty
\bibitem [{\citenamefont {Devereaux}\ and\ \citenamefont
  {Hackl}(2007)}]{devereaux2007inelastic}%
  \BibitemOpen
  \bibfield  {author} {\bibinfo {author} {\bibfnamefont {T.~P.}\ \bibnamefont
  {Devereaux}}\ and\ \bibinfo {author} {\bibfnamefont {R.}~\bibnamefont
  {Hackl}},\ }\href@noop {} {\bibfield  {journal} {\bibinfo  {journal} {Reviews
  of modern physics}\ }\textbf {\bibinfo {volume} {79}},\ \bibinfo {pages}
  {175} (\bibinfo {year} {2007})}\BibitemShut {NoStop}%
\bibitem [{\citenamefont {Hamm}\ and\ \citenamefont
  {Zanni}(2011)}]{hamm2011concepts}%
  \BibitemOpen
  \bibfield  {author} {\bibinfo {author} {\bibfnamefont {P.}~\bibnamefont
  {Hamm}}\ and\ \bibinfo {author} {\bibfnamefont {M.}~\bibnamefont {Zanni}},\
  }\href@noop {} {\emph {\bibinfo {title} {Concepts and methods of 2D infrared
  spectroscopy}}}\ (\bibinfo  {publisher} {Cambridge University Press},\
  \bibinfo {year} {2011})\BibitemShut {NoStop}%
\bibitem [{\citenamefont {Dorfman}\ \emph {et~al.}(2016)\citenamefont
  {Dorfman}, \citenamefont {Schlawin},\ and\ \citenamefont
  {Mukamel}}]{dorfman2016nonlinear}%
  \BibitemOpen
  \bibfield  {author} {\bibinfo {author} {\bibfnamefont {K.~E.}\ \bibnamefont
  {Dorfman}}, \bibinfo {author} {\bibfnamefont {F.}~\bibnamefont {Schlawin}},\
  and\ \bibinfo {author} {\bibfnamefont {S.}~\bibnamefont {Mukamel}},\
  }\href@noop {} {\bibfield  {journal} {\bibinfo  {journal} {Reviews of Modern
  Physics}\ }\textbf {\bibinfo {volume} {88}},\ \bibinfo {pages} {045008}
  (\bibinfo {year} {2016})}\BibitemShut {NoStop}%
\bibitem [{\citenamefont {Mukamel}(2000)}]{mukamel2000multidimensional}%
  \BibitemOpen
  \bibfield  {author} {\bibinfo {author} {\bibfnamefont {S.}~\bibnamefont
  {Mukamel}},\ }\href@noop {} {\bibfield  {journal} {\bibinfo  {journal}
  {Annual review of physical chemistry}\ }\textbf {\bibinfo {volume} {51}},\
  \bibinfo {pages} {691} (\bibinfo {year} {2000})}\BibitemShut {NoStop}%
\bibitem [{\citenamefont {Zhang}\ \emph {et~al.}(2007)\citenamefont {Zhang},
  \citenamefont {Kuznetsova}, \citenamefont {Meier}, \citenamefont {Li},
  \citenamefont {Mirin}, \citenamefont {Thomas},\ and\ \citenamefont
  {Cundiff}}]{zhang2007polarization}%
  \BibitemOpen
  \bibfield  {author} {\bibinfo {author} {\bibfnamefont {T.}~\bibnamefont
  {Zhang}}, \bibinfo {author} {\bibfnamefont {I.}~\bibnamefont {Kuznetsova}},
  \bibinfo {author} {\bibfnamefont {T.}~\bibnamefont {Meier}}, \bibinfo
  {author} {\bibfnamefont {X.}~\bibnamefont {Li}}, \bibinfo {author}
  {\bibfnamefont {R.~P.}\ \bibnamefont {Mirin}}, \bibinfo {author}
  {\bibfnamefont {P.}~\bibnamefont {Thomas}},\ and\ \bibinfo {author}
  {\bibfnamefont {S.~T.}\ \bibnamefont {Cundiff}},\ }\href@noop {} {\bibfield
  {journal} {\bibinfo  {journal} {Proceedings of the National Academy of
  Sciences}\ }\textbf {\bibinfo {volume} {104}},\ \bibinfo {pages} {14227}
  (\bibinfo {year} {2007})}\BibitemShut {NoStop}%
\bibitem [{\citenamefont {Cundiff}\ and\ \citenamefont
  {Mukamel}(2013)}]{cundiff2013optical}%
  \BibitemOpen
  \bibfield  {author} {\bibinfo {author} {\bibfnamefont {S.~T.}\ \bibnamefont
  {Cundiff}}\ and\ \bibinfo {author} {\bibfnamefont {S.}~\bibnamefont
  {Mukamel}},\ }\href@noop {} {\bibfield  {journal} {\bibinfo  {journal} {Phys.
  Today}\ }\textbf {\bibinfo {volume} {66}},\ \bibinfo {pages} {44} (\bibinfo
  {year} {2013})}\BibitemShut {NoStop}%
\bibitem [{\citenamefont {Lu}\ \emph {et~al.}(2017)\citenamefont {Lu},
  \citenamefont {Li}, \citenamefont {Hwang}, \citenamefont {Ofori-Okai},
  \citenamefont {Kurihara}, \citenamefont {Suemoto},\ and\ \citenamefont
  {Nelson}}]{lu2017coherent}%
  \BibitemOpen
  \bibfield  {author} {\bibinfo {author} {\bibfnamefont {J.}~\bibnamefont
  {Lu}}, \bibinfo {author} {\bibfnamefont {X.}~\bibnamefont {Li}}, \bibinfo
  {author} {\bibfnamefont {H.~Y.}\ \bibnamefont {Hwang}}, \bibinfo {author}
  {\bibfnamefont {B.~K.}\ \bibnamefont {Ofori-Okai}}, \bibinfo {author}
  {\bibfnamefont {T.}~\bibnamefont {Kurihara}}, \bibinfo {author}
  {\bibfnamefont {T.}~\bibnamefont {Suemoto}},\ and\ \bibinfo {author}
  {\bibfnamefont {K.~A.}\ \bibnamefont {Nelson}},\ }\href@noop {} {\bibfield
  {journal} {\bibinfo  {journal} {Physical review letters}\ }\textbf {\bibinfo
  {volume} {118}},\ \bibinfo {pages} {207204} (\bibinfo {year}
  {2017})}\BibitemShut {NoStop}%
\bibitem [{\citenamefont {Wan}\ and\ \citenamefont
  {Armitage}(2019)}]{wan2019resolving}%
  \BibitemOpen
  \bibfield  {author} {\bibinfo {author} {\bibfnamefont {Y.}~\bibnamefont
  {Wan}}\ and\ \bibinfo {author} {\bibfnamefont {N.}~\bibnamefont {Armitage}},\
  }\href@noop {} {\bibfield  {journal} {\bibinfo  {journal} {Physical review
  letters}\ }\textbf {\bibinfo {volume} {122}},\ \bibinfo {pages} {257401}
  (\bibinfo {year} {2019})}\BibitemShut {NoStop}%
\bibitem [{\citenamefont {Li}\ \emph {et~al.}(2021)\citenamefont {Li},
  \citenamefont {Oshikawa},\ and\ \citenamefont {Wan}}]{li2021photon}%
  \BibitemOpen
  \bibfield  {author} {\bibinfo {author} {\bibfnamefont {Z.-L.}\ \bibnamefont
  {Li}}, \bibinfo {author} {\bibfnamefont {M.}~\bibnamefont {Oshikawa}},\ and\
  \bibinfo {author} {\bibfnamefont {Y.}~\bibnamefont {Wan}},\ }\href@noop {}
  {\bibfield  {journal} {\bibinfo  {journal} {Physical Review X}\ }\textbf
  {\bibinfo {volume} {11}},\ \bibinfo {pages} {031035} (\bibinfo {year}
  {2021})}\BibitemShut {NoStop}%
\bibitem [{\citenamefont {Choi}\ \emph {et~al.}(2020)\citenamefont {Choi},
  \citenamefont {Lee},\ and\ \citenamefont {Kim}}]{choi2020theory}%
  \BibitemOpen
  \bibfield  {author} {\bibinfo {author} {\bibfnamefont {W.}~\bibnamefont
  {Choi}}, \bibinfo {author} {\bibfnamefont {K.~H.}\ \bibnamefont {Lee}},\ and\
  \bibinfo {author} {\bibfnamefont {Y.~B.}\ \bibnamefont {Kim}},\ }\href@noop
  {} {\bibfield  {journal} {\bibinfo  {journal} {Physical Review Letters}\
  }\textbf {\bibinfo {volume} {124}},\ \bibinfo {pages} {117205} (\bibinfo
  {year} {2020})}\BibitemShut {NoStop}%
\bibitem [{\citenamefont {Qiang}\ \emph {et~al.}(2023)\citenamefont {Qiang},
  \citenamefont {Quito}, \citenamefont {Trevisan},\ and\ \citenamefont
  {Orth}}]{qiang2023probing}%
  \BibitemOpen
  \bibfield  {author} {\bibinfo {author} {\bibfnamefont {Y.}~\bibnamefont
  {Qiang}}, \bibinfo {author} {\bibfnamefont {V.~L.}\ \bibnamefont {Quito}},
  \bibinfo {author} {\bibfnamefont {T.~V.}\ \bibnamefont {Trevisan}},\ and\
  \bibinfo {author} {\bibfnamefont {P.~P.}\ \bibnamefont {Orth}},\ }\href@noop
  {} {\bibfield  {journal} {\bibinfo  {journal} {arXiv preprint
  arXiv:2301.11243}\ } (\bibinfo {year} {2023})}\BibitemShut {NoStop}%
\bibitem [{\citenamefont {Hart}\ and\ \citenamefont
  {Nandkishore}(2022)}]{hart2022self}%
  \BibitemOpen
  \bibfield  {author} {\bibinfo {author} {\bibfnamefont {O.}~\bibnamefont
  {Hart}}\ and\ \bibinfo {author} {\bibfnamefont {R.}~\bibnamefont
  {Nandkishore}},\ }\href@noop {} {\bibfield  {journal} {\bibinfo  {journal}
  {arXiv preprint arXiv:2208.12817}\ } (\bibinfo {year} {2022})}\BibitemShut
  {NoStop}%
\bibitem [{\citenamefont {Fava}\ \emph {et~al.}(2022)\citenamefont {Fava},
  \citenamefont {Gopalakrishnan}, \citenamefont {Vasseur}, \citenamefont
  {Essler},\ and\ \citenamefont {Parameswaran}}]{fava2022divergent}%
  \BibitemOpen
  \bibfield  {author} {\bibinfo {author} {\bibfnamefont {M.}~\bibnamefont
  {Fava}}, \bibinfo {author} {\bibfnamefont {S.}~\bibnamefont
  {Gopalakrishnan}}, \bibinfo {author} {\bibfnamefont {R.}~\bibnamefont
  {Vasseur}}, \bibinfo {author} {\bibfnamefont {F.~H.}\ \bibnamefont
  {Essler}},\ and\ \bibinfo {author} {\bibfnamefont {S.}~\bibnamefont
  {Parameswaran}},\ }\href@noop {} {\bibfield  {journal} {\bibinfo  {journal}
  {arXiv preprint arXiv:2208.09490}\ } (\bibinfo {year} {2022})}\BibitemShut
  {NoStop}%
\bibitem [{\citenamefont {Gao}\ \emph {et~al.}(2023)\citenamefont {Gao},
  \citenamefont {Liu}, \citenamefont {Liao},\ and\ \citenamefont
  {Wan}}]{gao2023two}%
  \BibitemOpen
  \bibfield  {author} {\bibinfo {author} {\bibfnamefont {Q.}~\bibnamefont
  {Gao}}, \bibinfo {author} {\bibfnamefont {Y.}~\bibnamefont {Liu}}, \bibinfo
  {author} {\bibfnamefont {H.}~\bibnamefont {Liao}},\ and\ \bibinfo {author}
  {\bibfnamefont {Y.}~\bibnamefont {Wan}},\ }\href@noop {} {\bibfield
  {journal} {\bibinfo  {journal} {Physical Review B}\ }\textbf {\bibinfo
  {volume} {107}},\ \bibinfo {pages} {165121} (\bibinfo {year}
  {2023})}\BibitemShut {NoStop}%
\bibitem [{\citenamefont {Fava}\ \emph {et~al.}(2020)\citenamefont {Fava},
  \citenamefont {Coldea},\ and\ \citenamefont {Parameswaran}}]{fava2020glide}%
  \BibitemOpen
  \bibfield  {author} {\bibinfo {author} {\bibfnamefont {M.}~\bibnamefont
  {Fava}}, \bibinfo {author} {\bibfnamefont {R.}~\bibnamefont {Coldea}},\ and\
  \bibinfo {author} {\bibfnamefont {S.}~\bibnamefont {Parameswaran}},\
  }\href@noop {} {\bibfield  {journal} {\bibinfo  {journal} {Proceedings of the
  National Academy of Sciences}\ }\textbf {\bibinfo {volume} {117}},\ \bibinfo
  {pages} {25219} (\bibinfo {year} {2020})}\BibitemShut {NoStop}%
\bibitem [{\citenamefont {Morris}\ \emph {et~al.}(2021)\citenamefont {Morris},
  \citenamefont {Desai}, \citenamefont {Viirok}, \citenamefont {H{\"u}vonen},
  \citenamefont {Nagel}, \citenamefont {Room}, \citenamefont {Krizan},
  \citenamefont {Cava}, \citenamefont {McQueen}, \citenamefont {Koohpayeh}
  \emph {et~al.}}]{morris2021duality}%
  \BibitemOpen
  \bibfield  {author} {\bibinfo {author} {\bibfnamefont {C.}~\bibnamefont
  {Morris}}, \bibinfo {author} {\bibfnamefont {N.}~\bibnamefont {Desai}},
  \bibinfo {author} {\bibfnamefont {J.}~\bibnamefont {Viirok}}, \bibinfo
  {author} {\bibfnamefont {D.}~\bibnamefont {H{\"u}vonen}}, \bibinfo {author}
  {\bibfnamefont {U.}~\bibnamefont {Nagel}}, \bibinfo {author} {\bibfnamefont
  {T.}~\bibnamefont {Room}}, \bibinfo {author} {\bibfnamefont {J.}~\bibnamefont
  {Krizan}}, \bibinfo {author} {\bibfnamefont {R.}~\bibnamefont {Cava}},
  \bibinfo {author} {\bibfnamefont {T.}~\bibnamefont {McQueen}}, \bibinfo
  {author} {\bibfnamefont {S.}~\bibnamefont {Koohpayeh}}, \emph {et~al.},\
  }\href@noop {} {\bibfield  {journal} {\bibinfo  {journal} {Nature Physics}\
  ,\ \bibinfo {pages} {1}} (\bibinfo {year} {2021})}\BibitemShut {NoStop}%
\bibitem [{\citenamefont {Maartense}\ \emph {et~al.}(1977)\citenamefont
  {Maartense}, \citenamefont {Yaeger},\ and\ \citenamefont
  {Wanklyn}}]{maartense1977field}%
  \BibitemOpen
  \bibfield  {author} {\bibinfo {author} {\bibfnamefont {I.}~\bibnamefont
  {Maartense}}, \bibinfo {author} {\bibfnamefont {I.}~\bibnamefont {Yaeger}},\
  and\ \bibinfo {author} {\bibfnamefont {B.}~\bibnamefont {Wanklyn}},\
  }\href@noop {} {\bibfield  {journal} {\bibinfo  {journal} {Solid State
  Communications}\ }\textbf {\bibinfo {volume} {21}},\ \bibinfo {pages} {93}
  (\bibinfo {year} {1977})}\BibitemShut {NoStop}%
\bibitem [{\citenamefont {Kobayashi}\ \emph {et~al.}(1999)\citenamefont
  {Kobayashi}, \citenamefont {Mitsuda}, \citenamefont {Ishikawa}, \citenamefont
  {Miyatani},\ and\ \citenamefont {Kohn}}]{kobayashi1999three}%
  \BibitemOpen
  \bibfield  {author} {\bibinfo {author} {\bibfnamefont {S.}~\bibnamefont
  {Kobayashi}}, \bibinfo {author} {\bibfnamefont {S.}~\bibnamefont {Mitsuda}},
  \bibinfo {author} {\bibfnamefont {M.}~\bibnamefont {Ishikawa}}, \bibinfo
  {author} {\bibfnamefont {K.}~\bibnamefont {Miyatani}},\ and\ \bibinfo
  {author} {\bibfnamefont {K.}~\bibnamefont {Kohn}},\ }\href@noop {} {\bibfield
   {journal} {\bibinfo  {journal} {Physical Review B}\ }\textbf {\bibinfo
  {volume} {60}},\ \bibinfo {pages} {3331} (\bibinfo {year}
  {1999})}\BibitemShut {NoStop}%
\bibitem [{\citenamefont {Coldea}\ \emph {et~al.}(2010)\citenamefont {Coldea},
  \citenamefont {Tennant}, \citenamefont {Wheeler}, \citenamefont {Wawrzynska},
  \citenamefont {Prabhakaran}, \citenamefont {Telling}, \citenamefont
  {Habicht}, \citenamefont {Smeibidl},\ and\ \citenamefont
  {Kiefer}}]{coldea2010quantum}%
  \BibitemOpen
  \bibfield  {author} {\bibinfo {author} {\bibfnamefont {R.}~\bibnamefont
  {Coldea}}, \bibinfo {author} {\bibfnamefont {D.}~\bibnamefont {Tennant}},
  \bibinfo {author} {\bibfnamefont {E.}~\bibnamefont {Wheeler}}, \bibinfo
  {author} {\bibfnamefont {E.}~\bibnamefont {Wawrzynska}}, \bibinfo {author}
  {\bibfnamefont {D.}~\bibnamefont {Prabhakaran}}, \bibinfo {author}
  {\bibfnamefont {M.}~\bibnamefont {Telling}}, \bibinfo {author} {\bibfnamefont
  {K.}~\bibnamefont {Habicht}}, \bibinfo {author} {\bibfnamefont
  {P.}~\bibnamefont {Smeibidl}},\ and\ \bibinfo {author} {\bibfnamefont
  {K.}~\bibnamefont {Kiefer}},\ }\href@noop {} {\bibfield  {journal} {\bibinfo
  {journal} {Science}\ }\textbf {\bibinfo {volume} {327}},\ \bibinfo {pages}
  {177} (\bibinfo {year} {2010})}\BibitemShut {NoStop}%
\bibitem [{\citenamefont {Morris}\ \emph {et~al.}(2014)\citenamefont {Morris},
  \citenamefont {Aguilar}, \citenamefont {Ghosh}, \citenamefont {Koohpayeh},
  \citenamefont {Krizan}, \citenamefont {Cava}, \citenamefont {Tchernyshyov},
  \citenamefont {McQueen},\ and\ \citenamefont
  {Armitage}}]{morris2014hierarchy}%
  \BibitemOpen
  \bibfield  {author} {\bibinfo {author} {\bibfnamefont {C.}~\bibnamefont
  {Morris}}, \bibinfo {author} {\bibfnamefont {R.~V.}\ \bibnamefont {Aguilar}},
  \bibinfo {author} {\bibfnamefont {A.}~\bibnamefont {Ghosh}}, \bibinfo
  {author} {\bibfnamefont {S.}~\bibnamefont {Koohpayeh}}, \bibinfo {author}
  {\bibfnamefont {J.}~\bibnamefont {Krizan}}, \bibinfo {author} {\bibfnamefont
  {R.}~\bibnamefont {Cava}}, \bibinfo {author} {\bibfnamefont {O.}~\bibnamefont
  {Tchernyshyov}}, \bibinfo {author} {\bibfnamefont {T.}~\bibnamefont
  {McQueen}},\ and\ \bibinfo {author} {\bibfnamefont {N.}~\bibnamefont
  {Armitage}},\ }\href@noop {} {\bibfield  {journal} {\bibinfo  {journal}
  {Physical review letters}\ }\textbf {\bibinfo {volume} {112}},\ \bibinfo
  {pages} {137403} (\bibinfo {year} {2014})}\BibitemShut {NoStop}%
\bibitem [{\citenamefont {Kinross}\ \emph {et~al.}(2014)\citenamefont
  {Kinross}, \citenamefont {Fu}, \citenamefont {Munsie}, \citenamefont
  {Dabkowska}, \citenamefont {Luke}, \citenamefont {Sachdev},\ and\
  \citenamefont {Imai}}]{kinross2014evolution}%
  \BibitemOpen
  \bibfield  {author} {\bibinfo {author} {\bibfnamefont {A.}~\bibnamefont
  {Kinross}}, \bibinfo {author} {\bibfnamefont {M.}~\bibnamefont {Fu}},
  \bibinfo {author} {\bibfnamefont {T.}~\bibnamefont {Munsie}}, \bibinfo
  {author} {\bibfnamefont {H.}~\bibnamefont {Dabkowska}}, \bibinfo {author}
  {\bibfnamefont {G.}~\bibnamefont {Luke}}, \bibinfo {author} {\bibfnamefont
  {S.}~\bibnamefont {Sachdev}},\ and\ \bibinfo {author} {\bibfnamefont
  {T.}~\bibnamefont {Imai}},\ }\href@noop {} {\bibfield  {journal} {\bibinfo
  {journal} {Physical Review X}\ }\textbf {\bibinfo {volume} {4}},\ \bibinfo
  {pages} {031008} (\bibinfo {year} {2014})}\BibitemShut {NoStop}%
\bibitem [{\citenamefont {Kobayashi}\ \emph {et~al.}(2000)\citenamefont
  {Kobayashi}, \citenamefont {Mitsuda},\ and\ \citenamefont
  {Prokes}}]{kobayashi2000low}%
  \BibitemOpen
  \bibfield  {author} {\bibinfo {author} {\bibfnamefont {S.}~\bibnamefont
  {Kobayashi}}, \bibinfo {author} {\bibfnamefont {S.}~\bibnamefont {Mitsuda}},\
  and\ \bibinfo {author} {\bibfnamefont {K.}~\bibnamefont {Prokes}},\
  }\href@noop {} {\bibfield  {journal} {\bibinfo  {journal} {Physical Review
  B}\ }\textbf {\bibinfo {volume} {63}},\ \bibinfo {pages} {024415} (\bibinfo
  {year} {2000})}\BibitemShut {NoStop}%
\bibitem [{\citenamefont {White}(1992)}]{white1992density}%
  \BibitemOpen
  \bibfield  {author} {\bibinfo {author} {\bibfnamefont {S.~R.}\ \bibnamefont
  {White}},\ }\href@noop {} {\bibfield  {journal} {\bibinfo  {journal}
  {Physical review letters}\ }\textbf {\bibinfo {volume} {69}},\ \bibinfo
  {pages} {2863} (\bibinfo {year} {1992})}\BibitemShut {NoStop}%
\bibitem [{\citenamefont {Vidal}(2007)}]{vidal2007classical}%
  \BibitemOpen
  \bibfield  {author} {\bibinfo {author} {\bibfnamefont {G.}~\bibnamefont
  {Vidal}},\ }\href@noop {} {\bibfield  {journal} {\bibinfo  {journal}
  {Physical review letters}\ }\textbf {\bibinfo {volume} {98}},\ \bibinfo
  {pages} {070201} (\bibinfo {year} {2007})}\BibitemShut {NoStop}%
\bibitem [{\citenamefont {Sim}\ \emph {et~al.}(2023{\natexlab{a}})\citenamefont
  {Sim}, \citenamefont {Knolle},\ and\ \citenamefont
  {Pollmann}}]{sim2023nonlinear}%
  \BibitemOpen
  \bibfield  {author} {\bibinfo {author} {\bibfnamefont {G.}~\bibnamefont
  {Sim}}, \bibinfo {author} {\bibfnamefont {J.}~\bibnamefont {Knolle}},\ and\
  \bibinfo {author} {\bibfnamefont {F.}~\bibnamefont {Pollmann}},\ }\href@noop
  {} {\bibfield  {journal} {\bibinfo  {journal} {Physical Review B}\ }\textbf
  {\bibinfo {volume} {107}},\ \bibinfo {pages} {L100404} (\bibinfo {year}
  {2023}{\natexlab{a}})}\BibitemShut {NoStop}%
\bibitem [{\citenamefont {Kj{\"a}ll}\ \emph {et~al.}(2011)\citenamefont
  {Kj{\"a}ll}, \citenamefont {Pollmann},\ and\ \citenamefont
  {Moore}}]{kjall2011bound}%
  \BibitemOpen
  \bibfield  {author} {\bibinfo {author} {\bibfnamefont {J.~A.}\ \bibnamefont
  {Kj{\"a}ll}}, \bibinfo {author} {\bibfnamefont {F.}~\bibnamefont
  {Pollmann}},\ and\ \bibinfo {author} {\bibfnamefont {J.~E.}\ \bibnamefont
  {Moore}},\ }\href@noop {} {\bibfield  {journal} {\bibinfo  {journal}
  {Physical Review B}\ }\textbf {\bibinfo {volume} {83}},\ \bibinfo {pages}
  {020407} (\bibinfo {year} {2011})}\BibitemShut {NoStop}%
\bibitem [{\citenamefont {Robinson}\ \emph {et~al.}(2014)\citenamefont
  {Robinson}, \citenamefont {Essler}, \citenamefont {Cabrera},\ and\
  \citenamefont {Coldea}}]{robinson2014quasiparticle}%
  \BibitemOpen
  \bibfield  {author} {\bibinfo {author} {\bibfnamefont {N.~J.}\ \bibnamefont
  {Robinson}}, \bibinfo {author} {\bibfnamefont {F.~H.}\ \bibnamefont
  {Essler}}, \bibinfo {author} {\bibfnamefont {I.}~\bibnamefont {Cabrera}},\
  and\ \bibinfo {author} {\bibfnamefont {R.}~\bibnamefont {Coldea}},\
  }\href@noop {} {\bibfield  {journal} {\bibinfo  {journal} {Physical Review
  B}\ }\textbf {\bibinfo {volume} {90}},\ \bibinfo {pages} {174406} (\bibinfo
  {year} {2014})}\BibitemShut {NoStop}%
\bibitem [{\citenamefont {You}\ \emph {et~al.}(2016)\citenamefont {You},
  \citenamefont {Qiu},\ and\ \citenamefont {Ole{\'s}}}]{you2016quantum}%
  \BibitemOpen
  \bibfield  {author} {\bibinfo {author} {\bibfnamefont {W.-L.}\ \bibnamefont
  {You}}, \bibinfo {author} {\bibfnamefont {Y.-C.}\ \bibnamefont {Qiu}},\ and\
  \bibinfo {author} {\bibfnamefont {A.~M.}\ \bibnamefont {Ole{\'s}}},\
  }\href@noop {} {\bibfield  {journal} {\bibinfo  {journal} {Physical Review
  B}\ }\textbf {\bibinfo {volume} {93}},\ \bibinfo {pages} {214417} (\bibinfo
  {year} {2016})}\BibitemShut {NoStop}%
\bibitem [{\citenamefont {Laurell}\ \emph {et~al.}(2023)\citenamefont
  {Laurell}, \citenamefont {Alvarez},\ and\ \citenamefont
  {Dagotto}}]{laurell2023spin}%
  \BibitemOpen
  \bibfield  {author} {\bibinfo {author} {\bibfnamefont {P.}~\bibnamefont
  {Laurell}}, \bibinfo {author} {\bibfnamefont {G.}~\bibnamefont {Alvarez}},\
  and\ \bibinfo {author} {\bibfnamefont {E.}~\bibnamefont {Dagotto}},\
  }\href@noop {} {\bibfield  {journal} {\bibinfo  {journal} {Physical Review
  B}\ }\textbf {\bibinfo {volume} {107}},\ \bibinfo {pages} {104414} (\bibinfo
  {year} {2023})}\BibitemShut {NoStop}%
\bibitem [{\citenamefont {Paeckel}\ \emph {et~al.}(2019)\citenamefont
  {Paeckel}, \citenamefont {K{\"o}hler}, \citenamefont {Swoboda}, \citenamefont
  {Manmana}, \citenamefont {Schollw{\"o}ck},\ and\ \citenamefont
  {Hubig}}]{paeckel2019time}%
  \BibitemOpen
  \bibfield  {author} {\bibinfo {author} {\bibfnamefont {S.}~\bibnamefont
  {Paeckel}}, \bibinfo {author} {\bibfnamefont {T.}~\bibnamefont {K{\"o}hler}},
  \bibinfo {author} {\bibfnamefont {A.}~\bibnamefont {Swoboda}}, \bibinfo
  {author} {\bibfnamefont {S.~R.}\ \bibnamefont {Manmana}}, \bibinfo {author}
  {\bibfnamefont {U.}~\bibnamefont {Schollw{\"o}ck}},\ and\ \bibinfo {author}
  {\bibfnamefont {C.}~\bibnamefont {Hubig}},\ }\href@noop {} {\bibfield
  {journal} {\bibinfo  {journal} {Annals of Physics}\ }\textbf {\bibinfo
  {volume} {411}},\ \bibinfo {pages} {167998} (\bibinfo {year}
  {2019})}\BibitemShut {NoStop}%
\bibitem [{\citenamefont {Vidal}(2003)}]{vidal2003efficient}%
  \BibitemOpen
  \bibfield  {author} {\bibinfo {author} {\bibfnamefont {G.}~\bibnamefont
  {Vidal}},\ }\href@noop {} {\bibfield  {journal} {\bibinfo  {journal}
  {Physical review letters}\ }\textbf {\bibinfo {volume} {91}},\ \bibinfo
  {pages} {147902} (\bibinfo {year} {2003})}\BibitemShut {NoStop}%
\bibitem [{\citenamefont {White}\ and\ \citenamefont
  {Feiguin}(2004)}]{white2004real}%
  \BibitemOpen
  \bibfield  {author} {\bibinfo {author} {\bibfnamefont {S.~R.}\ \bibnamefont
  {White}}\ and\ \bibinfo {author} {\bibfnamefont {A.~E.}\ \bibnamefont
  {Feiguin}},\ }\href@noop {} {\bibfield  {journal} {\bibinfo  {journal}
  {Physical review letters}\ }\textbf {\bibinfo {volume} {93}},\ \bibinfo
  {pages} {076401} (\bibinfo {year} {2004})}\BibitemShut {NoStop}%
\bibitem [{\citenamefont {Hauschild}\ and\ \citenamefont
  {Pollmann}(2018)}]{hauschild2018efficient}%
  \BibitemOpen
  \bibfield  {author} {\bibinfo {author} {\bibfnamefont {J.}~\bibnamefont
  {Hauschild}}\ and\ \bibinfo {author} {\bibfnamefont {F.}~\bibnamefont
  {Pollmann}},\ }\href@noop {} {\bibfield  {journal} {\bibinfo  {journal}
  {SciPost Physics Lecture Notes}\ ,\ \bibinfo {pages} {005}} (\bibinfo {year}
  {2018})}\BibitemShut {NoStop}%
\bibitem [{\citenamefont {Sim}\ \emph {et~al.}(2023{\natexlab{b}})\citenamefont
  {Sim}, \citenamefont {Pollmann},\ and\ \citenamefont {Knolle}}]{sim_zenodo}%
  \BibitemOpen
  \bibfield  {author} {\bibinfo {author} {\bibfnamefont {G.}~\bibnamefont
  {Sim}}, \bibinfo {author} {\bibfnamefont {F.}~\bibnamefont {Pollmann}},\ and\
  \bibinfo {author} {\bibfnamefont {J.}~\bibnamefont {Knolle}},\ }\href
  {https://doi.org/10.5281/zenodo.7886700} {\bibinfo {title} {{Shedding Light
  on Microscopic Details: 2D Spectroscopy of 1D Quantum Ising Magnets}}}
  (\bibinfo {year} {2023}{\natexlab{b}})\BibitemShut {NoStop}%
\bibitem [{\citenamefont {Vidal}(2004)}]{vidal2004efficient}%
  \BibitemOpen
  \bibfield  {author} {\bibinfo {author} {\bibfnamefont {G.}~\bibnamefont
  {Vidal}},\ }\href@noop {} {\bibfield  {journal} {\bibinfo  {journal}
  {Physical review letters}\ }\textbf {\bibinfo {volume} {93}},\ \bibinfo
  {pages} {040502} (\bibinfo {year} {2004})}\BibitemShut {NoStop}%
\bibitem [{\citenamefont {White}\ and\ \citenamefont
  {Affleck}(2008)}]{white2008spectral}%
  \BibitemOpen
  \bibfield  {author} {\bibinfo {author} {\bibfnamefont {S.~R.}\ \bibnamefont
  {White}}\ and\ \bibinfo {author} {\bibfnamefont {I.}~\bibnamefont
  {Affleck}},\ }\href@noop {} {\bibfield  {journal} {\bibinfo  {journal}
  {Physical Review B}\ }\textbf {\bibinfo {volume} {77}},\ \bibinfo {pages}
  {134437} (\bibinfo {year} {2008})}\BibitemShut {NoStop}%
\bibitem [{\citenamefont {Verresen}\ \emph {et~al.}(2019)\citenamefont
  {Verresen}, \citenamefont {Moessner},\ and\ \citenamefont
  {Pollmann}}]{verresen2019avoided}%
  \BibitemOpen
  \bibfield  {author} {\bibinfo {author} {\bibfnamefont {R.}~\bibnamefont
  {Verresen}}, \bibinfo {author} {\bibfnamefont {R.}~\bibnamefont {Moessner}},\
  and\ \bibinfo {author} {\bibfnamefont {F.}~\bibnamefont {Pollmann}},\
  }\href@noop {} {\bibfield  {journal} {\bibinfo  {journal} {Nature Physics}\
  }\textbf {\bibinfo {volume} {15}},\ \bibinfo {pages} {750} (\bibinfo {year}
  {2019})}\BibitemShut {NoStop}%
\end{thebibliography}%
	
\end{document}